\documentclass[twocolumn,shwpacs,prc]{revtex4-1}
\usepackage{amssymb}
\usepackage[intlimits]{amsmath}
\usepackage{graphicx}
\usepackage{dcolumn}
\usepackage{amsfonts}
\usepackage{bm}
\begin{document}

\title{A systematic description of evaporation spectra for light and heavy  
compound nuclei.}
\author{R. J. Charity}
\affiliation{Department of Chemistry$,$ Washington University, St. Louis, Missouri
63130, USA}
\date{\today}

\begin{abstract}

To systematically describe evaporation spectra for light and heavy compound nuclei over a large 
range of excitation energies, it was necessary to consider three ingredients in the statistical model. 
Firstly, transmission coefficients or barrier penetration factors for charged-particle emission
are typically taken from global fits to elastic-scattering data. However, 
such transmission coefficients do not reproduce the barrier region of evaporation 
spectra and reproduction of the data requires a distributions of Coulomb barriers. This is 
possibly associated with large fluctuations in the compound-nucleus shape or density profile. 
Secondly for heavy nuclei, an excitation-energy dependent level-density parameter is 
required to describe the slope of the exponential tails of these spectra. The level-density 
parameter was reduced at larger temperatures, consistent with the expected 
fadeout of long-range correlation, but the strong $A$ dependence of this effect is unexpected.
Lastly to describe the angular-momentum dependence of the level density in light nuclei at large spins,
the macroscopic rotational energy of the nucleus has to be reduced from the values predicted 
with the Finite-Range Liquid-Drop model.  
\end{abstract}

\maketitle

 \section{INTRODUCTION}
The statistical model of compound-nucleus (CN) decay is  extensively used in
pure and applied nuclear science. In many reaction scenarios one or more 
compound nuclei are formed after a nuclear collisions. Compound nuclei are 
equilibrated in their non-decay degrees of freedom and thus their decay is 
independent of how they were created. Statistical-model codes have been 
used as ``afterburners'' in many reaction-modelling programs decaying 
the simulated compound nuclei produced from some initial fast reaction 
mechanism. The initial reaction could be fusion, spallation, 
fragmentation, etc. 

Accurate determination of the statistical-model parameters at high 
excitation energies would give insight into properties of hot nucleus.
The level densities are sensitive to the magnitude of long-range 
correlations associated with collective excitations, 
transmission coefficients are sensitive to the charge and mass distributions 
and fission maybe sensitive to the nuclear viscosity. Efforts to extract 
such information require systematic studies of compound-nucleus decay covering
a large range of compound-nucleus $Z$,$A$ and excitation energy.

The modelling of spallation reactions is important in applications ranging 
from transmutation of nuclear waste, the design of neutron sources for 
condensed-mater studies, radiation protection around accelerators and in 
space, and for the production of rare isotopes for nuclear and astrophysics 
experiments. The modelling of such reactions involves an 
Intra-Nuclear Cascade  \cite{Yariv81,Boudard02} or Quantum Molecular Dynamics  code \cite {ou09} to 
simulate the production of the initial fast reaction products and the properties 
of the residual compound nuclei formed ($Z$, $A$, $E^*$, and $J$ joint distributions). 
These compound nuclei then de-excited with a statistical-model code which includes 
evaporation and fission and possible other decay modes. The residual nuclei are predicted to 
be excited to large excitation energies (many hundreds of MeV) and therefore knowledge of the 
statistical-model parameters is needed for this energy regime. 

The final predictions of spallation modelling are sensitive to both the 
statistical-model parameters and those associated with the 
initial fast phase of the reaction. When fitting experimental spallation 
data, it is not always possible to isolate the role of the statistical-model 
parameters and constrain them. Alternatively heavy-ion-induced
complete-fusion reactions can be used to create compound nuclei. 
In complete fusion, the 
excitation energy and identity of the compound nucleus are completely defined
from conservation laws.
The CN spin distribution can also be well constrained. The maximum spin 
can be determined from measurements of the 
total fusion cross section or, alternatively, simple one-dimensional models 
are generally quite accurate above the fusion barrier. 
Thus the simple complete-fusion mechanism with no fast 
non-statistical particles and a well defined distribution of CN provides 
an opportunity to constrain the statistical-model parameters. 

Of course complete-fusion reactions are limited by preequilibrium emissions 
and incomplete-fusion processes which sets it at large bombarding energies ($>$ 10 MeV/$A$).
However large excitation energies (up to $\sim$250 MeV) can still be probed with 
complete fusion using more symmetric reactions. Heavy-ion-induced fusion reactions, 
especially the more symmetric cases, emphasize large spins, typically larger 
than those probed by spallation reactions at the same excitation energies. 
Therefore application of statistical-model parameters determined in fusion reactions
to spallation modelling requires a good understanding the  spin dependence of 
CN decay.

The statistical model has a long history in heavy-ion induced fusion reactions and has 
been fit to a large body of data including fission probabilities, light-particle evaporation spectra, 
residual $Z$ and $A$ distributions, gamma-ray multiplicities, etc. Although such data are usually 
fit within the statistical-mode framework, it has generally been found necessary 
to fine tune the statistical-model parameters for a particular compound nucleus or mass region. No
statistical-model prescription exists which gives accurate predictions of these quantities 
over the entire table of isotopes. This work starts to address these problems by concentrating on 
light-particle evaporation which is sensitive to the excitation energy and spin dependences of the 
nuclear level density and the transmission coefficients for penetration of the Coulomb barriers 
hindering particle emission.

The assumption that the decay of the compound nucleus is independent of
how its was created may not always be correct in fusion reactions. 
At high excitation energies when the statistical lifetime approaches the 
fusion timescales, dynamical effects may occur which depend on the entrance-channel 
mass asymmetry. Specifically symmetric reaction channels are predicted to 
dissipate the entrance-channel kinetic energy more slowly and may start particle 
evaporation before the fusion dynamics is complete. There have been many studies
of entrance-channel dependence of compound-nucleus decay.
However taken as a whole, no clear consistent picture has emerged from these studies 
and in a number of cases their conclusions are contradictory. In particular concerning
the shapes of evaporation spectra, one should note three studies where $\alpha$-particle
spectra  were measured for different entrance channels, but with matched 
excitation-energy and spin distributions. Cinausero \textit{et al.} found no entrance-channel
dependence of the spectral shape for $A\sim$160 compound nuclei at $E^*\sim$300~MeV formed in 
$^{86}$Kr,+$^{76}$Ge, $^{16}$O+$^{150}$Sm, and $^{60}$Ni+$^{100}$Mo reactions \cite{Cinausero96}.
For $E^*$=170~MeV $^{164}$Yb compound nuclei 
formed in $^{16}$O+$^{148}$Sm and $^{64}$Ni+$^{100}$Mo 
reactions, Charity \textit{et al.} noted a slight enhancement in the $\alpha$-particle yield
in the subbarrier region, otherwise the kinetic-energy spectra were consistent \cite{Charity97}.
On the other hand, Liang \textit{et al.} reported on entrance-channel dependences of the slope of 
the high-energy tail in $E^*$=113-MeV $^{156}$Er compound nuclei formed in $^{12}$C+$^{144}$Sm,
$^{35}$Cl+$^{121}$Sb, and $^{60}$Ni+$^{96}$Mo reactions \cite{Liang97}. It is difficult to 
reconcile there three studies as they pertain to the same mass region. 
In this work we will ignore such effects and assume, that if they exist, they 
are small at least compared to the overall variations due to the 
mass, excitation-energy, and spin dependences of the statistical-model parameters 

Statistical-model parameters are extracted from comparison of statistical-model
calculations to experimental data. In this work, all statistical-model calculations
were performed with the code {\sc GEMINI++} \cite{Charity08} written in the C++ language. 
This is a successor of the well known statistical-model code {\sc GEMINI} \cite{Charity88a} 
written in FORTRAN.

\section{DATA}
\label{Sec:data}

The data used in this work to constrain the statistical-model parameters 
has come from many 
experimental studies covering a wide range of compound-nucleus masses. 
The compound nucleus, the reactions, the excitation energies and 
references are listed in Table~\ref{Tbl:CN}. For compound nucleus with 
$A>150$, only studies where light particles were detected in coincidence with 
evaporation residues were used. For the lighter systems, only inclusive
spectra are available. By appropriate selection of detection angle 
[backward (forward) angles for normal (reverse) kinematics reactions], 
one can isolate proton and $\alpha$-particle spectra which are dominated by 
compound-nucleus emission though some contamination from other reaction 
processes is possible for the lowest kinetic energies \cite{Komarov07}.
This will be discussed in more detail in Sec.~\ref{Sec:light}.

While the residue-gated spectra may be cleaner, they may suffer from 
distortions due to the limited kinematic acceptance of the residue detectors. 
For example, detection of evaporation residues at large angles enhances 
high-energy particles as these give the largest recoil kick to the residue 
enabling it to get to such angles. The spectra used in this work were 
either corrected for this effect in referenced studies, or, for the $^{160}$Yb 
compound nucleus, the {\sc GEMINI++} simulations were gated on the 
experimental residue acceptance.

It is important in the statistical-model calculations to have realistic spin
distributions for the compound nuclei. The fusion cross section as a 
function of spin was assumed to have the form:
\begin{equation}
\sigma_{fus}(J) = \pi \lambdabar^2 \sum \frac{(2J+1)}
{1+\exp\left(\frac{J-J_0}{\delta J}\right)} 
\label{Eq:fus}
\end{equation}
The quantity $J_0$ can be constrained from the fusion cross section. This 
is either measured, constrained from systematics, or obtained from 
the Bass model \cite{Bass74,*Bass77} which is reasonably accurate for 
the systems under study. The parameter $\delta J$ was varied from 
2 to 10~$\hbar$ with increasing asymmetry of the entrance channel. However in this work, 
the sensitivity of the predicted evaporated spectra to this parameter is very 
small.

Fission competition is also important for determining the $J$ values which 
give rise to evaporation residues. When available (Table~\ref{Tbl:CN}), 
fission and/or evaporation residue cross sections were fit by  adjusting 
the fission parameter $a_{f}/a_{n}$ of Sec.~\ref{Sec:fission}. Otherwise,
interpolated values of $a_{f}/a_{n}$ were used. A more detailed discussion
of the fission parameters in {\sc GEMINI++} can be found in 
Ref.~\cite{Mancusi10}.

\begin{table*}[tbp]
\caption{Experimental data used in this work indicating the compound nucleus (CN),
the beam energy $E_{beam}$, the excitation energy $E^*$, the fusion reaction,
the evaporation spectra measured (\textit{n},\textit{p},$\alpha$), the values of $J_0$
defining the angular-momentum distribution of Eq.~(\ref{Eq:fus}). The first listed reference
refer to the study that measured the kinetic energy spectra. 
The $\sigma$ references refers to measurements of the fission and residues cross sections
used to constrain $J_0$ and the fission probability.}
\label{Tbl:CN}%
\begin{ruledtabular}
\begin{tabular}{cccD{+}{+}{4}cccc}
 CN       & $E_{beam}$& $E^*$&   \multicolumn{1}{c}{reaction}   &      ref.     & spectra & $\sigma$ refs. &  $J_0$ \\
    & [MeV] & [MeV] &              &  &           \\
\hline
$^{59}$Cu & 100 & 58   & ^{32}\mbox{S}+^{27}\mbox{Al}        & \cite{Fornal88}   & $\alpha$   & \cite{Gutbrod73,Kozub75,Puhlhofer77,Rosner85}         & 27\footnotemark[1]    \\
          & 105 & 60   &                                     & \cite{Choudhury84}& $\alpha$   & \cite{Gutbrod73,Kozub75,Puhlhofer77,Rosner85}         & 30\footnotemark[1]     \\
          & 130 & 72   &                                     & \cite{Fornal88}   & $\alpha$   & \cite{Gutbrod73,Kozub75,Puhlhofer77,Rosner85}         & 34\footnotemark[1]     \\
          & 140 & 77   &                                     & \cite{Fornal88}   & $\alpha$   & \cite{Gutbrod73,Kozub75,Puhlhofer77,Rosner85}         & 38\footnotemark[1]    \\
          & 150 & 82   &                                     & \cite{Fornal88}   & $\alpha$   & \cite{Gutbrod73,Kozub75,Puhlhofer77,Rosner85}         & 39\footnotemark[1]    \\
          & 214 & 110  &                                     & \cite{Choudhury84}& $\alpha$   & \cite{Gutbrod73,Kozub75,Puhlhofer77,Rosner85}         & 45\footnotemark[1]     \\
$^{67}$Ga & 187 & 90   & ^{40}\mbox{Ar}+^{27}\mbox{Al}       & \cite{LaRana87}   & \textit{p},$\alpha$ & & 46\footnotemark[2] \\
          & 670 & 127  & ^{55}\mbox{Mn}+^{12}\mbox{C}        & \cite{Brown99}    & \textit{p},$\alpha$ & & 42\footnotemark[2] \\
          & 280 & 127  & ^{40}\mbox{Ar}+^{27}\mbox{Al}       & \cite{Brown99}    & \textit{p},$\alpha$ & & 54\footnotemark[2] \\
$^{96}$Ru & 180 & 113  & ^{32}\mbox{S}+^{64}\mbox{Ni}        & \cite{Kildir92}   & \textit{p},$\alpha$ & & 69\footnotemark[2]   \\
$^{106}$Cd&160  &99    &^{32}\mbox{S}+^{74}\mbox{Ge}         &\cite{Nebbia94}     & \textit{p},$\alpha$ & & 68\footnotemark[3]  \\  
          &99   &291   &                                     &\cite{Nebbia94}     & \textit{p},$\alpha$ & & 83\footnotemark[3]  \\
          &99   &291   &                                     &\cite{Nebbia94}     & \textit{p},$\alpha$ & & 89\footnotemark[3]  \\
          &99   &291   &                                     &\cite{Nebbia94}     & \textit{p},$\alpha$ & & 89\footnotemark[3]   \\ 
$^{117}$Te&81   &71    &^{14}\mbox{N}+^{103}\mbox{Rh}        &\cite{Galin74}      & \textit{p},$\alpha$ & & 40\footnotemark[3]\\
          & 146 & 71   &^{40}\mbox{Ar}+^{77}\mbox{Se}        &\cite{Galin74a}     & \textit{p},$\alpha$ & & 52\footnotemark[3] \\
          &121  &106   &^{14}\mbox{N}+^{103}\mbox{Rh}        &\cite{Galin74}      & \textit{p},$\alpha$ & & 53\footnotemark[3]\\
$^{156}$Er& 142 & 113  &^{12}\mbox{C}+^{144}\mbox{Sm}        &\cite{Liang97}      & \textit{p},$\alpha$ & \cite{Liang97} & 54\footnotemark[1]                  \\
          & 218 & 113  &^{35}\mbox{Cl}+^{121}\mbox{Sb}       &\cite{Liang97}      & \textit{p}          & & $>$86\footnotemark[4]  \\
          & 333 & 113  &^{60}\mbox{Ni}+^{96}\mbox{Zr}        &\cite{Liang97}      & \textit{p},$\alpha$ &\cite{Janssens86} & $>$90\footnotemark[4]  \\     
$^{160}$Yb& 300  &91   &^{60}\mbox{Ni}+^{100}\mbox{Mo}       &\cite{Charity03}    &  \textit{n},\textit{p},$\alpha$& \cite{Charity03} &$>$90\footnotemark[4]\\
          & 360  &129   &                                     &\cite{Charity03}    &  \textit{n},\textit{p},$\alpha$ &\cite{Charity03}&$>$90\footnotemark[4] \\
          & 420  &166   &                                     &\cite{Charity03}    &   \textit{n},\textit{p},$\alpha$ &\cite{Charity03}&$>$90\footnotemark[4] \\ 
          & 480  &204   &                                     &\cite{Charity03}    & \textit{n},\textit{p},$\alpha$&\cite{Charity03}&$>$90\footnotemark[4] \\
          & 546  &245   &                                     &\cite{Charity03}    &  \textit{n},\textit{p},$\alpha$&\cite{Charity03}&$>90$\footnotemark[4] \\
$^{193}$Tl& 145 &65    &^{28}\mbox{Si}+^{160}\mbox{Ho}       &\cite{Fineman94}    &  \textit{p},$\alpha$  & \cite{Fineman94}&46\footnotemark[1]\\
          & 166 &83    &                                     &\cite{Fineman94}    &  \textit{p},$\alpha$  & \cite{Fineman94}&68\footnotemark[1]\\
          & 193 &106   &                                     &\cite{Fineman94}    &  \textit{p},$\alpha$  & \cite{Fineman94}&80\footnotemark[1]\\
          & 216 &125   &                                     &\cite{Fineman94}    &  \textit{p},$\alpha$  & \cite{Fineman94} &96\footnotemark[1]\\
$^{200}$Pb& 121  &86   &^{19}\mbox{F}+^{181}\mbox{Ta}        &\cite{Caraley00}    &  \textit{p},$\alpha$  & \cite{Caraley00}&56\footnotemark[1] \\
          & 154  &116   &                                     & \cite{Caraley00}   &  \textit{p},$\alpha$  & \cite{Caraley00}&72\footnotemark[1] \\
          & 179  &139   &                                     & \cite{Caraley00}   &  \textit{p},$\alpha$  & \cite{Caraley00}&83\footnotemark[1]\\
          & 195  &153   &                                     & \cite{Caraley00}   &  \textit{p},$\alpha$  & \cite{Caraley00}&89\footnotemark[1]\\
$^{224}$Th& 114 &59    &^{16}\mbox{O}+^{208}\mbox{Pb}        & \cite{Fineman94}   &  \textit{p},$\alpha$&\cite{Videbaek77,Brinkmann94,Back85}  & 51\footnotemark[1] \\
$^{224}$Th& 138&82     &                                     & \cite{Fineman94}   &  \textit{p},$\alpha$&\cite{Videbaek77,Brinkmann94,Back85}  &67\footnotemark[1]\\
\end{tabular}
\end{ruledtabular}
\footnotetext[1] {The $J_0$ is constrained from the measured fusion cross section.}
\footnotetext[2] {The $J_0$ values were obtained from the referenced studied where the fusion cross section was estimated from the 
systematics of Ref.~\cite{Frobrich84}.}
\footnotetext[3] {The $J_0$ values were estimated from the Bass model \cite{Bass74,*Bass77}.}
\footnotetext[4] {The $J_0$ vlaues are large and the residue cross section is determined solely by fission competition. Fission parameters were adjusted to
reproduce measured evaporation residues.}

\end{table*}

\section{EVAPORATION FORMALISM}
\label{Sec:evap}

As {\sc GEMINI++} is to be used for CN with high spins, 
the evaporation of light particles is treated with the Hauser-Feshbach 
formalism \cite{Hauser52} which explicitly takes into account the 
spin degrees of freedom.
The partial decay width of
a compound nucleus of excitation energy $E^{\ast }$ and spin $J_{CN}$ for
the evaporation of particle $i$ is

\begin{multline}
\Gamma _{i}(E^*,J_{CN})=\frac{1}{2\pi \rho _{CN}\left( E^{\ast },J_{CN}\right) }\int
d\varepsilon \sum_{J_{d}=0}^{\infty }
\\
\sum_{J=\left\vert
J_{CN}-J_{d}\right\vert }^{J_{CN}+J_{d}}\sum_{l=\left\vert 
J-S_{i}\right\vert }^{J+S_{i}}
T_{\ell }\left( \varepsilon \right) \rho_d
\left( E^{\ast }-B_{i}-\varepsilon ,J_{d}\right)
\end{multline}%
where $J_{d}$ is the spin of the daughter nucleus, $S_{i}$, $J$, and $\ell $,
are the spin, total and orbital angular momenta of the evaporated particle, 
$\varepsilon $ and $B_{i}$ are is its kinetic and separation energies, $%
T_{\ell }$ is its transmission coefficient or barrier penetration factor,
and $\rho_d$ and $\rho _{CN}$ are the level densities of the daughter and
compound nucleus, respectively. The summations include all angular momentum
couplings between the initial and final states. In {\sc GEMINI++}, the 
Hauser-Feshbach formalism is implemented for the \textit{n}, \textit{p}, \textit{d} , 
\textit{t}, $^{3}$He, $\alpha $, $^{6}$He, $^{6-8}$Li, and $^{7-10}$Be channels. However
in this work, we will just compare predicted kinetic-energy spectra  
to experimental results for the \textit{p}, $\alpha$, and occasional \textit{n} channels.
{\sc GEMINI++} also allows for intermediate-mass fragment emission follow the formalism 
of Moretto \cite{Moretto75}. However, these decay modes are not very important for calculations 
of this work.

The nuclear level density is often approximated by the Fermi-gas form \cite{Bohr75}  
derived for a spherical nucleus in the independent-particle model with 
constant single-particle level densities;
\begin{gather}
\rho_{FG}\left( E^{\ast } ,J\right) = 
\frac{ (2J+1)}{24\sqrt{2}\, a^{1/4}\, U^{5/4} 
\, \sigma^{3} }
 \exp(S) \, , \\
S = 2 \sqrt{a U}
\label{eq:fermiGas}
\end{gather}
where $S$ is the nuclear entropy and the level-density parameter is
\begin{equation}
a = \frac{\pi^2}{6} \left[ g^n(\varepsilon^n_F) + g^p(\varepsilon^p_F)\right].
\label{eq:littleA}
\end{equation}
Here $g^n(\varepsilon^n_F)$ and $g^p(\varepsilon^p_F)$ are the neutron and proton
single-particle level densities  at their respective Fermi energies and
\begin{gather}
U = E^* - E_{rot}(J),\quad E_{rot} = \frac{J(J+1) \hbar^2 }{2 \mathcal{I}_{rig}}, \\
\sigma^2 = \mathcal{I}_{rig} T.
\label{Eq:Erot}
\end{gather}
The quantity $\mathcal{I}_{rig}$ is the moment of inertia of a rigid body with 
the same density
distribution as the nucleus and $T$ is the nuclear temperature; 
\begin{equation}
\frac{1}{T} = \frac{dS}{dU}
\label{Eq:temp}
\end{equation}
The quantity $U$ can be interpreted as a thermal 
excitation, after the rotational energy of the nucleus is removed.

At large angular momenta, macroscopic models of the nucleus such as the 
Rotating Liquid-Drop Model (RLDM) \cite{Cohen74} and Sierk's Yukawa-plus-exponential finite-range 
calculations \cite{Sierk86} predict the nuclear shape distorts to accommodate the centrifugal 
forces. Many implementations of the statistical model including {\sc GEMINI++}, 
generalize  
Eq.(\ref{Eq:Erot}) by the replacing $E_{rot}(J)$, the rotational energy of a spherical 
nucleus of fixed moment of inertia, with $E_{yarst}(J)$, the deformation-plus-rotational energy 
predicted by these macroscopic models where the deformation increased with spin. In {\sc GEMINI++},
the Sierk predictions of $E_{yrast}(J)$ are used for all but the lightest compound nuclei
(see Sec.~\ref{Sec:light}).

The shape of the kinetic-energy spectra of evaporated particle is thus 
sensitive to three ingredients.
\begin{itemize}
\item The magnitude of the level-density parameter and its excitation-energy
dependence
\item The transmission coefficients $T_{\ell }\left( \varepsilon \right)$.
\item The angular-momentum dependence of $E_{yrast}(J)$.
\end{itemize}

The level-density parameter defines the slope of the exponential tail of 
the evaporation spectrum while the transmission coefficients define the shape 
in the Coulomb barrier region and the effects of these two ingredients 
are easily isolated when comparing to data. The angular-momentum 
dependence of $E_{yrast}(J)$ is most important in light nuclei where the moments 
of inertia are small and thus  $E_{yrast}(J)$ rises rapidly with spin. 
In particular $E_{yrast}$ has a strong influence on the heavier fragments such as 
$\alpha$ particles which can remove large amounts spins. For these particles, 
the functional form of $E_{yrast}(J)$ can make significant modifications to 
the predicted shape of the evaporation spectrum in the exponential tail and even in the 
Coulomb barrier region. The effect of $E_{yrast}$ can be disentangled from the effects
of the level-density parameter and the transmission coefficients by comparing data for
a lighter particle such as a proton to that for a heavier particle such as an $\alpha$ 
particle.

In the following three sections, the parametrization of these three ingredients needed to
describe experiment data will be described. We will discuss light and heavy systems
separately.  

\section{HEAVY COMPOUND NUCLEI}
\begin{figure}[tbp]
\includegraphics*[scale=0.4]{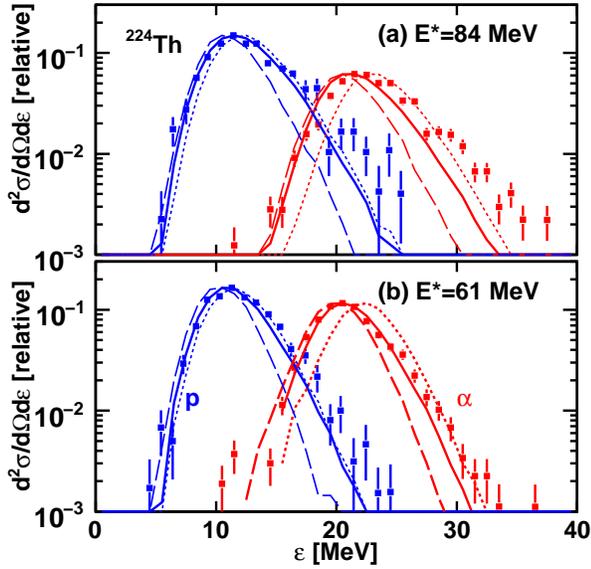}
\caption{(Color online) Center-of-mass kinetic-energy spectra 
of $\alpha$ particles and protons 
detected in coincidence with evaporation residues formed in  
$^{16}$O+$^{208}$Pb reactions. Experimental results (data points) 
are shown for the indicated excitation energies of the $^{224}$Th compound 
nuclei. The curves show spectra predicted with {\sc GEMINI++} code and 
normalized to the same peak height as the experimental data. The solid curves
 (the default calculations of the code)
were obtained with the excitation-dependent level-density parameter  
and with distributions of Coulomb barriers. The short-dashed curves indicated 
the results obtained using a single Coulomb barrier and the long-dashed curves are 
associated with an excitation-independent $\widetilde{a}$=A/7.3~MeV$^{-1}$ 
level-density parameter.} 
\label{fig:th224}
\end{figure}

\begin{figure}[tbp]
\includegraphics*[scale=0.4]{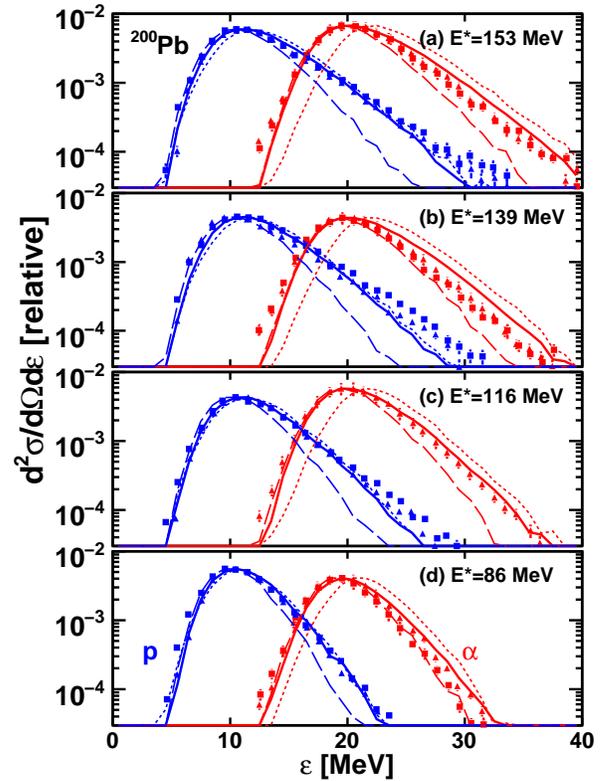}
\caption{(Color online) As in Fig.~\ref{fig:th224}
but now for $^{200}$Pb compound nuclei formed in $^{19}$F+$^{181}$Ta 
reactions.}
\label{fig:pb200}
\end{figure}

\begin{figure}[tbp]
\includegraphics*[scale=0.4]{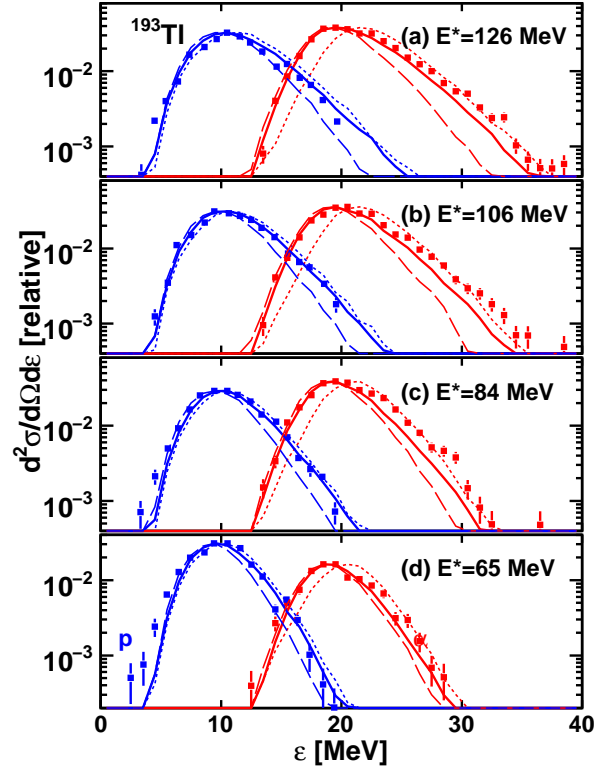}
\caption{(Color online) As in Fig.~\ref{fig:th224}
but now for $^{193}$Tl compound nuclei formed in $^{32}$Si+$^{160}$Ho 
reactions.}
\label{fig:tl193}
\end{figure}

\begin{figure}[tbp]
\includegraphics*[scale=0.5]{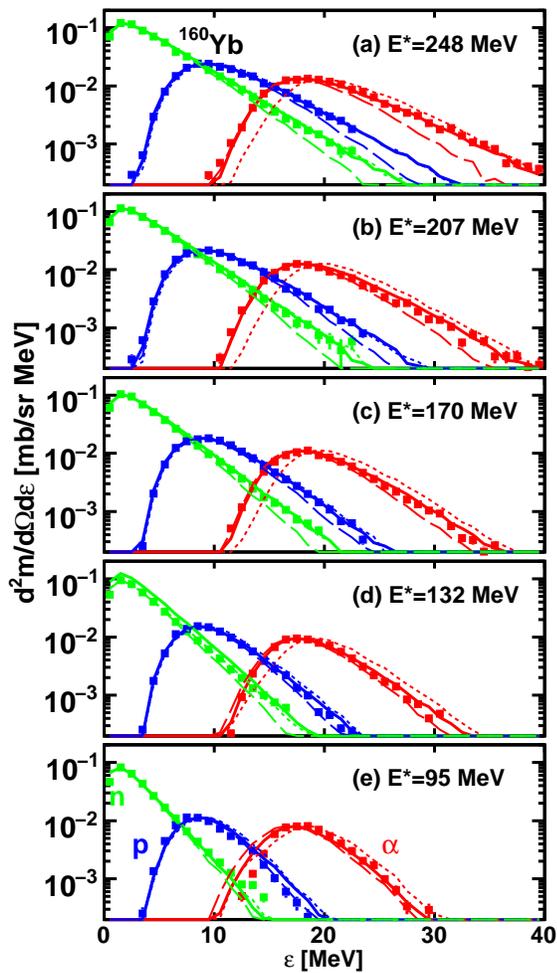}
\caption{(Color online) As in Fig.~\ref{fig:th224}
but now for $^{160}$Yb compound nuclei formed in $^{60}$Ni+$^{100}$Mo 
reactions with neutron spectra also included.}
\label{fig:yb160}
\end{figure}

\begin{figure}[tbp]
\includegraphics*[scale=0.4]{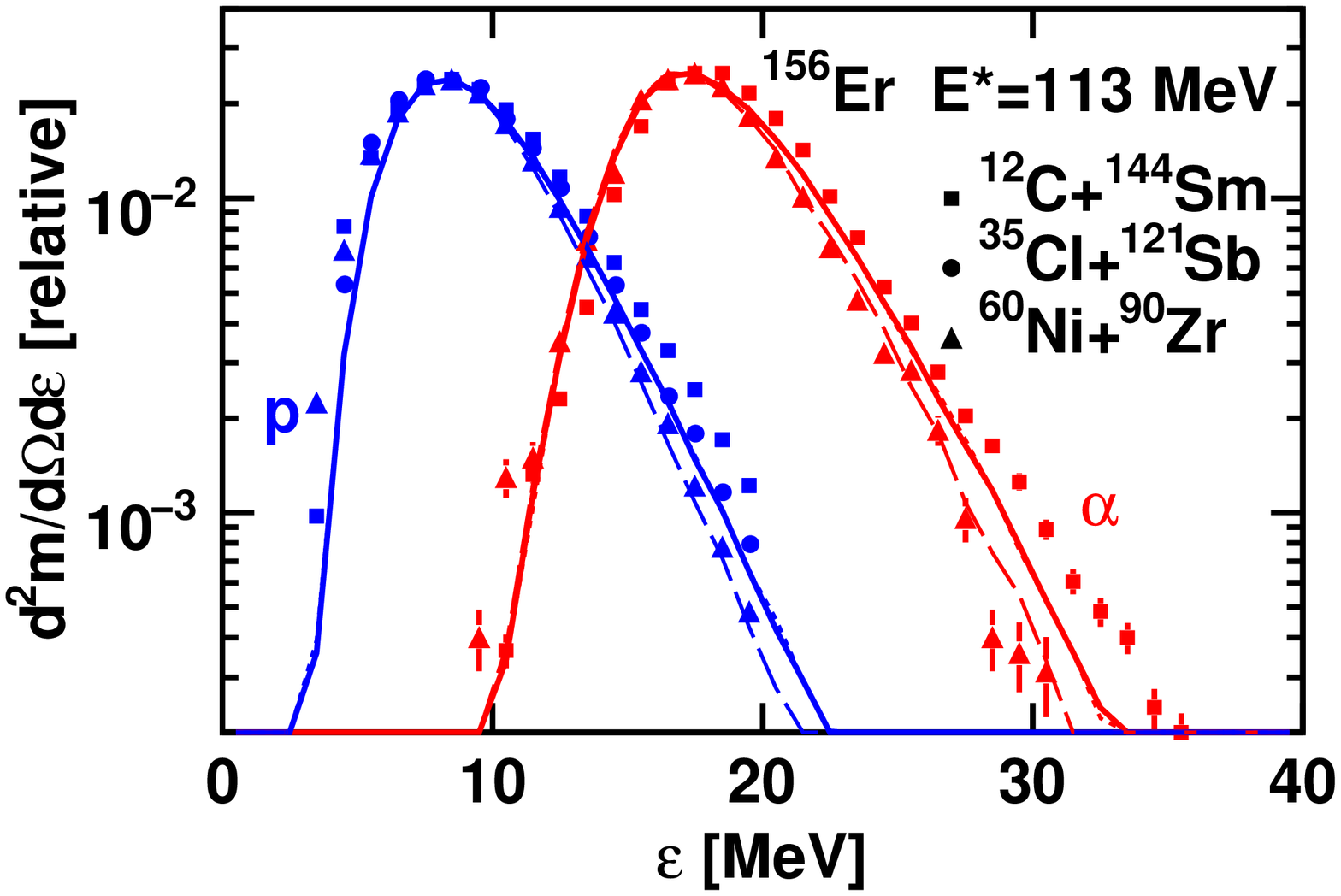}
\caption{(Color online) As in Fig.~\ref{fig:th224}
but now for $^{156}$Er compound nuclei formed in 
the three indicated reactions.}
\label{fig:er156}
\end{figure}

Let us start by concentrating on the heavier compound nuclei with 
$A>$150 for which the evaporation spectra are shown in 
Figs.~\ref{fig:th224} to 
\ref{fig:er156}. These data sets were all obtained with a  coincidence 
requirement of a detected evaporation residue. We will first consider which 
transmission coefficients and level-densities allow us to reproduced the 
shape of the experimental spectra. Predicted spectra in these figures 
will be normalized to give the same peak differential multiplicity in order 
to concentrate of the reproduction the spectral shapes. Subsequently we will
return to consider how well one can reproduce the absolute multiplicities
of evaporated protons and $\alpha$ particles.

\subsection{Transmission Coefficients}

The evaporation formalism is justified on the condition of detailed balance. 
The evaporation rate of an isolated compound nucleus is assumed identical 
to the emission rate of such a nucleus in equilibrium with a gas of the 
evaporated particles. In equilibrium there is a balance between the 
emission and the inverse, absorption rates of that particle and thus the 
transmission coefficients or barrier penetration probabilities should be 
identical to those for  the inverse absorption process.

Transmission coefficients have traditionally been obtained from the inverse
reaction using  optical-model parameters obtained from global
optical-model fits to elastic-scattering data. There are two problems 
with this approach. 
First, Alexander \textit{et al. }\cite{Alexander90} have
pointed out that such transmission coefficients contains the effects of
transparency in the inverse reaction which is not appropriate in
evaporation. Instead is was suggested that the real optical-model potentials 
should still be used, but
to ensure full absorption, the incoming-wave boundary-condition (IWBC) model 
\cite{Rawitscher66} be used to calculate $T_{\ell }$. In {\sc GEMINI++}, 
global optical-model
potentials were obtained from Refs.~\cite%
{Perey62,Perey63,Wilmore64,McFadden66,Becchetti71,Cook82,Balzer77}.
The difference between IWBC and optical-model transmission coefficients
is only important for neutrons, protons, deuterons, tritons, and $^3$He particles
 as other particles experience 
strong absorption inside the Coulomb barrier. 
Due to transparency, optical-model transmission 
coefficients for nucleons do not approach unity for energies well 
above the barrier as is the case the for IWBC values. However, the difference
between IWBC and standard optical-model $T_{\ell }$ values is not that large 
and it is difficult to differentiate them based on experimental data
due to uncertainties in other statistical-model parameters.  Comparisons 
of statistical-model predictions with IBWC and standard optical-model value of 
$T_{\ell}$  are made in Ref.~\cite{Kildir92,Kildir95} where the biggest differences 
are associated with deuteron and triton spectra;

The more important problem with the traditional transmission 
coefficients is that they are not associated with the inverse reaction. 
The true inverse process to evaporation is the absorption of the particle by
a hot, rotating target nuclei which is impossible to measure experimentally.
This is highlighted by the fact that IWBC and optical-model transmission 
coefficients fail to reproduce the shape of the low-energy or ``sub-barrier'' 
region of the spectra of $\alpha$ and other heavier particles \cite%
{Nicolis90,Gonin90,Kildir92,Boger94,Fineman94,Charity97,Liang97,Caraley00}.
We illustrate this in Figs. \ref{fig:th224} to \ref{fig:er156} where 
statistical-model predictions obtained 
with {\sc GEMINI++} using the IWBC transmission coefficients, 
indicated by the short-dashed curves, are compared to experimental data. The 
level-density prescription used in these calculations
will be described in the following sections and Sierk's values of $E_{yrast}$
were used.  
For $\alpha$ particles emitted from these heavier systems, the relative yield
in the ``sub-barrier'' region is clearly underpredicted.

\begin{figure}[tbp]
\includegraphics*[scale=0.4]{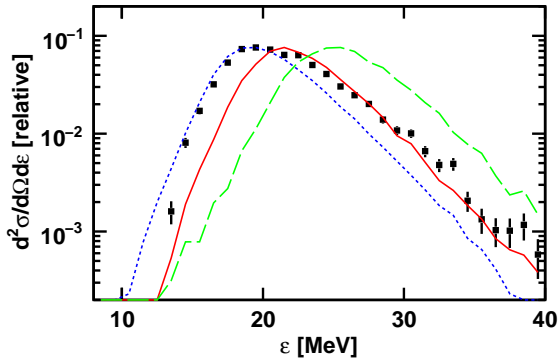}
\caption{(Color online) Comparison of the experimental $\alpha$-particle
evaporation spectrum (data points) measured in the $^{28}$Si+$^{160}$Ho
reaction producing $^{193}$Tl compound nuclei at $E^*$=126~MeV to 
{\sc GEMINI++} predictions. The solid curve was obtained with standard
 IWBC transmission coefficients, while the short- and long-dashed curves 
were obtained by increasing and decreasing the radius parameter of the 
nuclear potential, respectively. (see text). The curves have been normalized 
to the same peak height as the experimental data. }  
\label{fig:tl193Bar}
\end{figure}

Some studies have attempted to reproduce such data by reducing 
the Coulomb barrier, for example by allowing an extended radial-profile of a 
spherical nucleus \cite{Lacey87}. 
However, a simple reduction in the barrier, just shifts the kinetic-energy 
spectrum down in energy. The experimental $\alpha$-particle spectra have more 
rounded maxima than predictions 
with such barriers. This is illustrated in Fig.~\ref{fig:tl193Bar} where
 the $\alpha$ spectrum measured for $E^*$=120~MeV $^{193}$Tl compound 
nuclei formed in 
$^{28}$Si+$^{160}$Ho reactions is compared to a number of calculations. 
The solid curve is again the prediction with the standard 
IWBC transmission coefficients. 
For the short-dashed curve, the Coulomb barrier was decreased by 
increasing the radius parameter of the nuclear potential by $\delta r$
from its original value of $R_{0}$ in the global optical-model potential. 
The value of $\delta r$ is temperature dependent and is given later.
With the reduced barrier, there is a predicted increased in the yield at 
lower energies but the yield starts dropping too early with energy and doesn't
reproduce the width of the experimental distribution. For interest sake, the 
spectrum predicted with $=R_{0}-\delta r$ is indicated by the long-dashed 
curve. Although decreasing the
 level-density parameter will increase the predicted width of the spectrum, the exponential 
slope of the experimental spectrum is already reproduced for $E_{c.m.}>$27~MeV by all
the curves.  It is clear that if one considered a distribution of radius parameters, 
one could increase the predicted width of the $\alpha$-particle spectrum.
This conclusion was  also found for evaporated Li and Be 
particles \cite{Charity01}. 

A distribution could arise from a static nuclear deformation if 
evaporation is averaged over the nuclear surface \cite{Huizenga89}. 
Alternatively, the
origin of this distribution may have contributions from compound-nucleus
thermal shape fluctuations \cite{Charity00,Charity01a} and/or fluctuation in
the diffuseness of the nuclear surface or nuclear size.

\begin{figure}[tbp]
\includegraphics*[scale=0.4]{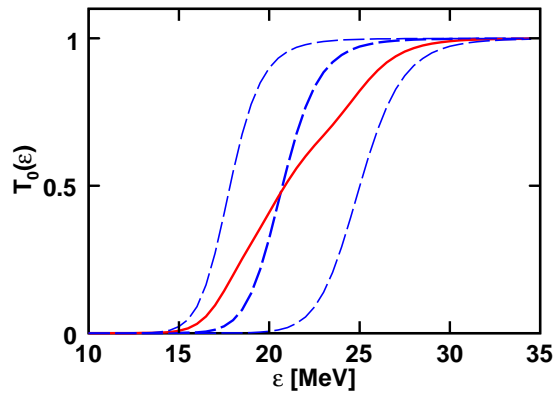}
\caption{(Color online) Transmission coefficents for $\alpha$+$^{193}$Tl at 
$\ell$=0. The dashed curves shows the three transmission coefficents 
which different nuclear radii which are averaged in Eq.~(\ref{Eq:trans}) 
and the sold curve is the result.}
\label{fig:trans}
\end{figure}

If the fluctuations are thermally
induced then we expect, to first order,  their variance to be proportional to temperature. 
In {\sc GEMINI++}, a simple scheme was implemented to 
incorporate the effects of barrier distributions. 
The transmission coefficients were calculated as 
\begin{equation}
T_{\ell }\left( \varepsilon \right) =\frac{T_{\ell }^{R_{0}-\delta r}\left(
\varepsilon \right) +T_{\ell }^{R_{0}}\left( \varepsilon \right) +T_{\ell
}^{R_{0}+\delta r}\left( \varepsilon \right) }{3}
\label{Eq:trans}
\end{equation}
which is the average of three IWBC transmission coefficients calculated with
three different radius parameters of the nuclear potential. 
It was assumed
\begin{equation} 
\delta r=w \sqrt{T}
\end{equation}
consistent with thermal fluctuations where the value of the
parameter $w=$1.0~fm was obtained from fits to experiment data and 
$T$ is the nuclear temperature of the daughter nucleus as defined 
in Eq.~(\ref{Eq:temp}). An example of these transmission coefficients 
is shown in Fig.~\ref{fig:trans} for $\alpha$+$^{193}$Tl with $\ell=0$ at 
T=3~MeV. The dashed curves show three transmission coefficients associated 
with the three radii in Eq.~(\ref{Eq:trans}) and the solid curve is the final 
result, the average of the three dashed curves. The more gradual 
rise of the transmission with kinetic energy gives rise to a broader peak 
in the predicted $\alpha$-particle spectra.

Results obtained with this prescription are indicated by the solid curves in 
Figs.~\ref{fig:th224} to \ref{fig:er156} 
and generally reproduce the $\alpha$ particle data quite well. 

Because of their lower absolute Coulomb barriers, the effect of the distribution is 
much less for protons and is practically absent for neutrons. However, 
the agreement for protons is generally improved.

One should note that the magnitudes of the fluctuations are very large. 
For a temperature of $T$=3~MeV, $\delta r$ is $\sim$ 25\% of the nuclear
radius for $A$=160. For ellipsoidal shape fluctuations in 
Ref.~\cite{Charity01}, the full width at half maximum of the Coulomb 
barrier distributions was predicted to be only $\sim$7\%. 
This suggests that either higher-order shape fluctuations are required or 
the fluctuations are associated with density profile. 

The effects of the barrier distributions is to increase the width of the 
kinetic-energy window around the barrier where the transmission coefficients 
change significantly. For example in Fig.~\ref{fig:trans}, the 
transmission coefficient changed from 10\% to 90\% over an interval of 4.5~MeV 
for IWBC calculation [$T_{\ell}^{R_0}(\epsilon)$]. However, with 
Eq.~(\ref{Eq:trans}), this increased to 9.2~MeV.   
An alternative way of increasing the width of this 
window would be to make the radial width of the barrier narrower. 
Narrow barriers allow for more 
tunneling and enhance the transmission just below the barrier and also 
decrease it just above the barrier. However it is difficult to see how 
the barrier could be made significantly narrower as the decrease in the 
potential at large distances is dictated by the Coulomb potential 
which falls off slowly. Thus barrier distributions are the most likely 
explanation.    

\subsection{Level-Density Parameter}

The slope of the exponential tail of the kinetic-energy spectrum gives
sensitivity to the nuclear temperature $T$ [Eq.~(\ref{Eq:temp})]. 
The temperature is dependent on the rate of change of the level density,
but not its absolute value.

The Fermi-gas level density prescription of Sec.~\ref{Sec:evap} 
can be further refined by including the pairing interaction \cite{Santo63,Moretto72}.
For the spin and excitation-energy region of interest in this work,
the pairing gap has vanished and we can use a back-shifted Fermi-Gas formula
by substituting the following definition of the thermal excitation energy
\begin{equation}
U = E^* - E_{yrast}(J) + \delta P 
\end{equation}
where $\delta P$ is the pairing correction to the empirical mass formula.

\begin{figure}[tbp]
\includegraphics*[scale=0.45]{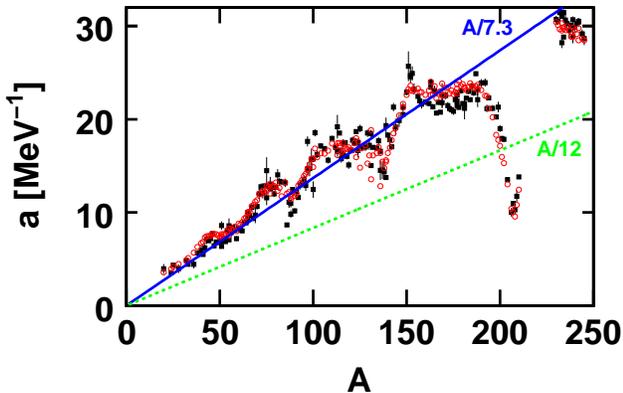}
\caption{(Color online) Mass dependence of level-density parameters.
Experimental points from neutron-resonance counting are shown as the
filled, square data points. The open circles are fits obtained using Eq.~%
(\protect\ref{eq:Ignatyuk}).}
\label{fig:aden}
\end{figure}

At low excitation energies, the absolute level density
can be measured via neutron-resonance counting. The level-density parameters
extracted from such data in Ref.~ \cite{John01}, using the back-shifted  
Fermi-gas formula, are plotted in Fig.~\ref{fig:aden}.
The level-density parameter has strong fluctuations due to shell effects which
can be parametrized as \cite{Ignatyuk75};%
\begin{equation}
a\left( U\right) =\widetilde{a}\left[ 1-h\left( U/\eta + J/J_{\eta} \right) 
\frac{\delta W%
}{U}\right]  \label{eq:Ignatyuk}
\end{equation}%
where $\delta W$ is the shell correction to the liquid-drop mass and $%
\widetilde{a}$ is a smoothed level-density parameter. With $h(x)=\tanh (x)$ 
we obtain a best fit (open-circular points) to the experimental data 
with $\eta $=19~MeV and $\widetilde{a}=A$/7.3~MeV$^{-1}$. 

The angular-momentum dependence of $h(x)$ is irrelevant for neutron resonances which 
are S-wave in nature. However for fusion reactions, it was decided to include a fading out of 
shell effects which spin. Although at high spins and low values of $U$, shell 
corrections are still important, the configuration of the nucleus has changed from
the ground state and the use of the ground-state shell correction is wrong. Rather 
than use an incorrect shell correction, it was decided to use no correction at all.
The parameter $J_{\eta}$ was set to 50 $\hbar$.
  
The above prescription for the fadeout of shell and pairing corrections
is used in all {\sc GEMINI++} calculations with separation energies $B_{i}$, 
nuclear masses, shell $\delta W$ and pairing $\delta P$ corrections obtained 
from the tabulations of M\"{o}ller \textit{et al.}~\cite{Moller95}. 

Predicted kinetic-energy spectra obtained using these pairing and shell modified
Fermi-gas level density prescription are shown as the long-dashed curves in 
Figs.~\ref{fig:th224} to ~\ref{fig:er156}. They significantly
underestimate the yield in the exponential tails for the heavier systems.
This disagreement gets worse with both increasing compound-nucleus mass and
excitation energy. These results suggest that a excitation-dependent
value of $\widetilde{a}$ is needed. 

The  value of the smoothed level-density parameter $\widetilde{a}$
used in these calculations is large compared to 
estimates from the independent-particle model of $\widetilde{a}=$
$A$/10-$A$/11 MeV$^{-1}$ \cite{Ignatyuk76,charity05} and the 
difference has been
attributed to correlations. In particular, it is the long-range correlations
associated with coupling of nucleon single-particle degrees of freedom to
low-lying collective modes and giant resonances which are most important.

It has been proposed that long-range correlations modify the Fermi-gas level
density in two ways. The first of these is called collective enhancement 
\cite{Bjornholm74,Hansen83}. For example if we have a deformed nucleus, then
for each single-particle configuration, one can consider collective
rotations. In additions, both spherical and deformed nuclei can have
collective vibrational motions. These collective motions give rise to
rotational and vibrational bands enhancing the level density above the
single-particle value, i.e., 
\begin{equation}
\rho(E^*) = K_{coll}(E^*) \rho_{FG}(E^*)
\end{equation}
where $K_{coll}$ is the collective enhancement factor.

Long-range correlations, and to a lesser extent also short-range
correlations, cause an enhancement of the single-particle level densities 
$g^{n}(\varepsilon^n_{F})$ and $g^{p}(\varepsilon^p_{F})$ in Eq.~(\ref{eq:littleA})
\cite{Mahaux91} which leads to an enhancement in $a$.
This enhancement is counterbalanced by the effect of nonlocality. In
fact without the correlations, we would expect smaller level-density
parameters than the predicted $\widetilde{a}=$ $A$/10-$A$/11~MeV$^{-1}$ 
values due to
the unbalanced effect of nonlocality. As $U$ increases, long-range
correlations are expected to wash out giving rise to both a disappearance of
collective enhancement ($K_{coll}\rightarrow1$) 
and a reduction in the level-density parameter itself
\cite{Bjornholm74,Hansen83,Shlomo91}.

In this work, we interpret level densities through the Fermi-gas formula, i.e.,
take Eq.~(\ref{eq:fermiGas}) as correct by definition, but use an effective
level-density parameter $\widetilde{a}_{eff}$ that is enhanced above the 
single-particle estimate of Eq.~(\ref{eq:littleA}) and decreases with excitation
energy due to the fade out of these long-range correlations, i.e.,
\begin{equation}
\rho(E^*) = \rho_{FG}(E^*,\widetilde{a}_{eff}) = K_{coll}(E^*) \rho_{FG}(E^*,\widetilde{a}).
\end{equation}
At low energies, $\widetilde{a}_{eff}$ is set to the value of $A$/7.3~MeV$^{-1}$ to be 
consistent with the counting of neutron resonances. 

We have parametrized its excitation-energy dependence by
\begin{equation}
\widetilde{a}_{eff}\left( U\right) =\frac{A}{k_{\infty }-(k_{\infty }-k_{0})\exp
\left( -\frac{\kappa }{k_{\infty }-k_{0}}\frac{U}{A}\right) }
\label{eq:fita}
\end{equation}%
where $k_{0}$=7.3~MeV and the asymptotic value at high excitation energy is $%
\widetilde{a}_{eff}$=$A/k_{\infty }$. The parameter $\kappa $ defines how fast the
long-range correlations wash out with excitation energy. This expression is
expected to be valid only to moderately high excitation energies where 
expansion and increases in the surface diffuseness \cite{Shlomo91,Sobotka06} 
are not significant.

Experimental evidence for an excitation-energy dependence of 
$\widetilde{a}_{eff}$
was found in the $A\sim $160 region; measurements of light-particle
evaporation spectra (\textit{n}, \textit{p}, $\alpha $) with excitation
energies ranging from 50 to 250~MeV \cite{Charity03,Komarov07} show clear
evidence of a departure from a constant value of $\widetilde{a}_{eff}$ with the data
being reproduced by the parametrization 
\begin{equation}
\widetilde{a}_{eff}\left( U\right) =\frac{A}{k_{0}+\kappa U/A}
\end{equation}%
when $k_{0}$=7 MeV and $\kappa $=1.3 MeV. This equation is just a
lower-order approximation of Eq.~(\ref{eq:fita}). From an examination of other
studies on evaporation spectra, it is apparent that there is a strong $A$
dependence of $\kappa $. Nebbia \textit{et al.} \cite{Nebbia94}
find no deviation from a constant $\widetilde{a}_{eff}$ value for the $^{106}$Cd
CN with excitation energies up to 291 MeV \cite{Nebbia94}. Whereas for
heavier systems, larger values of $\kappa $ are deduced; values of $\kappa $%
=2-3 were found for $A\sim $200 ($E^{\ast }<$150~MeV) \cite%
{Fineman94,Caraley00} and $\kappa $=8.5 for $A$=224 ($E^{\ast }<$90) \cite%
{Fineman94} with $k_{0}$=8 MeV.

\begin{figure}[tbp]
\includegraphics*[scale=0.4]{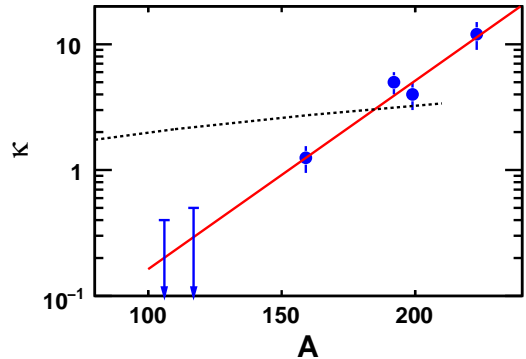}
\caption{(Color online) Values of $\protect\kappa$ in 
Eq.~(\protect\ref{eq:fita}) obtained from fitting evaporation spectra. 
The solid line shows a 
smooth
approximation used to calculate evaporation spectra and ER excitation functions. 
The dashed curve shows $\kappa$ values extracted from the predictions of Ref.~\cite{Shlomo91}.}
\label{fig:kappa}
\end{figure}

In this work, we have made a systematic study of the $A$ dependence 
of $\kappa $ by fitting the evaporation spectra with Eq.~(\ref{eq:fita}). 
At the excitation energies studied,
we cannot constrain the value of $k_{\infty }$ and it was set to
12~MeV. The fitted values of $\kappa $ obtained from reproducing the
evaporation spectra in Figs.~\ref{fig:th224} to \ref{fig:yb160} 
are plotted verses $A$ in Fig.~\ref{fig:kappa}. For a 
single compound nucleus, the values of $\kappa$ obtained from fitting the 
proton and $\alpha$-spectra were similar though not always identical and  
the error bars in  Fig.~\ref{fig:kappa} reflect this range of $\kappa$ values.

In addition to these data points, Fig.~\ref{fig:kappa} gives some limits 
for $\kappa$ obtained from $^{117}$Te and $^{106}$Cd compound nuclei.
These data are in fact consistent with $\kappa$=0 and will be discussed in 
more detail in Sec.~\ref{Sec:light}.
 
Figure~\ref{fig:kappa} is a log plot and it indicates that $\kappa $ 
increases very rapidly with mass number. Although we do not have enough 
data points to determine this dependence in detail, we have fitted it with the
exponential function shown by the solid line in this figure and given by
\begin{equation}
\kappa (A) = 0.00517 \exp(0.0345 A).
\end{equation}

\begin{figure}[tbp]
\includegraphics*[scale=0.4]{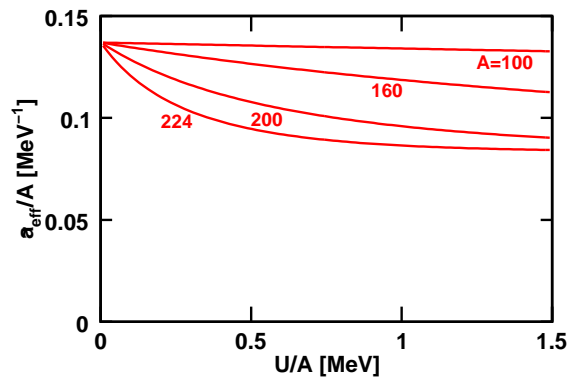}
\caption{(Color online) Excitation-energy dependence of the smoothed 
level-density parameter obtained in this work for the indicated $A$ values. }
\label{fig:littleA}
\end{figure}

The excitation-energy dependence of the level-density parameter associated
with this dependence is illustrated in Fig.~\ref{fig:littleA} for the 
indicated $A$ values. The excitation dependence is very strong for the 
heaviest compound nuclei, but below $A<$100, there is very little dependence.  
Statistical-model calculations preformed with this dependence are indicated 
by the solid curves in Figs.~\ref{fig:th224} to \ref{fig:er156}. They 
reproduce the data much better than a constant 
$\widetilde{a}_{eff}$=$A$/7.3~MeV$^{-1}$, 
though they are not perfect. 

The individual fits to each reactions (not shown) 
are slightly better but quite similar. In the similar mass $^{193}$Tl and 
$^{200}$Pb systems, the tails of the $\alpha$-particles spectra are under and 
over predicted, respectively. This could just be an artifact 
due to small experimental errors in the two studies or may reflect an asymmetry
$(N-Z)/A$ dependence of $\kappa$ or even an entrance channel effect.
Also the $^{224}$Th data clearly suffer from large statistical errors due to 
the very small residue cross sections.  
Further systematic measurements of a large number of compound nuclei with the
same experimental apparatus would help resolve these issues.  

More sophisticated calculations of nuclear level density have been obtained 
within the Shell-Model Monte Carlo method but only for light nuclei such as $^{56}$Fe 
have calculations been extended to  high
excitation energies \cite{Alhassid03}. These calculated level densities can be fit with a 
constant level-density parameter of value $A$/9.5~MeV$^{-1}$. This basically 
consistent with the results of this work in that the level density of 
light nuclei has a Fermi-gas form ($\widetilde{a}_eff$ independent of $U$), 
however the value of 
$A$/9.5~MeV$^{-1}$ is a little smaller than the value 
$A$/7.3~MeV$^{-1}$ used in this work.

\subsection{Multiplicities and Cross Sections}
\begin{figure}[tbp]
\includegraphics*[scale=0.4]{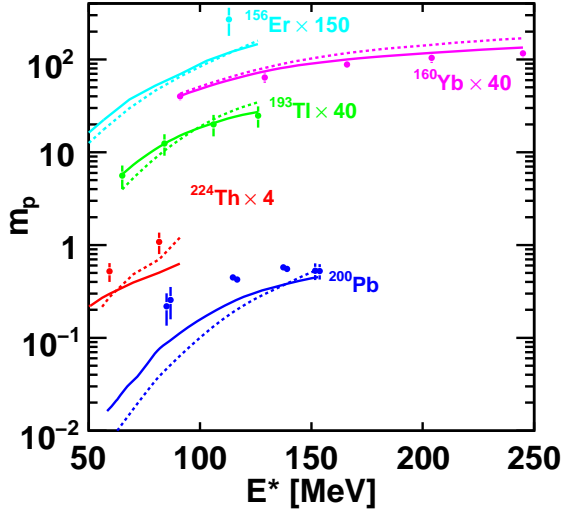}
\caption{(Color online) Comparison of experimental and predicted proton 
multiplicities from the indicated compound nuclei. To aid in viewing, 
the data have been scaled by the indicated factors. The solid curves were 
obtained with the excitation-dependent level-density parameter 
and with distribution of Coulomb barriers. The dashed
curves shows the prediction with single Coulomb barriers and a constant 
$\widetilde{a}_{eff}=A/7.3$~MeV$^{-1}$.}
\label{fig:multProton}
\end{figure}

\begin{figure}[tbp]
\includegraphics*[scale=0.4]{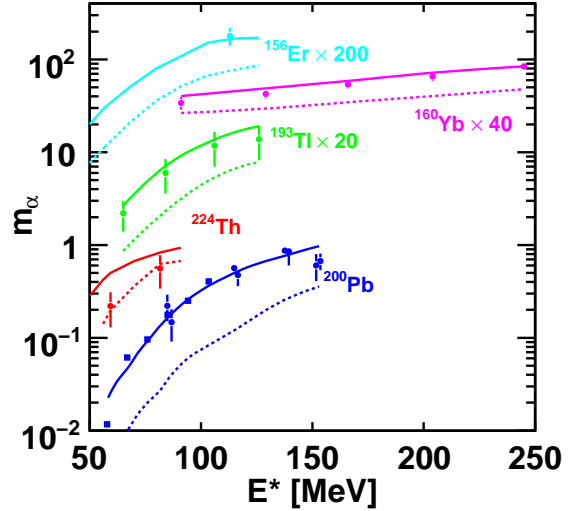}
\caption{(Color online) As for Fig.~\ref{fig:multProton} but now for 
$\alpha$-particle multiplicities.}
\label{fig:multAlpha}
\end{figure}

So far we have only considered the shapes of the kinetic energy spectra.
It is also important to determined the accuracy to which the absolute yields of
evaporated particles can be predicted. For the $^{156}$Er, $^{160}$Yb, $^{193}$Tl,
$^{200}$Pb, and $^{224}$Th compound nuclei for which light particles were detected in 
coincidence with evaporation residues, the predicted multiplicities 
are compared to the experimental proton and 
$\alpha$-particles values in Figs.~\ref{fig:multProton} and \ref{fig:multAlpha}.
To separate the data from the different systems, the multiplicities were scaled by the 
indicated amounts. The solid curves in both figures show calculations with 
the default setting of the code, i.e., distribution of Coulomb barriers and 
an excitation-dependent level-density parameter $\widetilde{a}_{eff}$.
They reproduce the $\alpha$-particle data quite well. For protons, 
the $^{160}$Yb and $^{193}$Tl data are well reproduced, while the other systems
underpredict the multiplicities by up to a factor of 2.

It is difficult to understand how a better overall reproduction of the experimental proton
multiplicities can be obtained for $A~\sim$160. For example the $^{156}$Er and $^{160}$Yb compound nuclei
have similar $Z$ and $A$ values, are both produced in Ni induced reactions and thus explore similar 
spin distributions. The protons are predicted to be emitted at large excitation energies 
where shell and pairing effects are expected to be washed out. Modifications to {\sc GEMINI++} 
that increase the proton multiplicity for the $^{156}$Er system will 
also increase the multiplicities for the $^{160}$Yb system in disagreement with the experimental
data. One should consider whether the inability to simultaneously fits these two systems 
is an experimental problem.

These predicted multiplicities are quite sensitive to the level-density and Coulomb
barrier prescription. To illustrate this, the dashed curves in Fig.~\ref{fig:multProton} and
\ref{fig:multAlpha} were obtained with a constant 
$\widetilde{a}_{eff} = A/7.3$~MeV$^{-1}$ 
and with the IWBC transmission coefficients for a single Coulomb barrier. 
For $\alpha$ particles, this results in a large decrease
of the multiplicities by a factor of 3 to 10. Clearly the level density 
and Coulomb barrier distribution are important to correctly predict these multiplicities. 
For protons we are somewhat less sensitive to these ingredients.  

\subsection{Consequence for Fission}
\label{Sec:fission}

Although this work is not focused on the fission probability, it is 
interesting to determine the consequences of the parametrizations in the 
preceding sections on the fission probability.
Fission was first incorporated into the statistical model by 
Bohr and Wheeler using the transition-state formalism first introduced 
to calculate chemical reaction rates.   
The Bohr-Wheeler decay width \cite {Bohr39} is
\begin{equation}
\Gamma_{BW}(E^*,J) = \frac{\pi}{\rho_{CN}(E^*,J)}\int \rho_s(E^*-B_f(J)-\epsilon,J) 
d\epsilon
\end{equation}
where $B_f(J)$ is the spin-dependent fission barrier, and $\rho_S$ is the level density at the transition state, i.e.,
the saddle-point configuration. The variable $\epsilon$ is the kinetic energy in the fission 
degree of freedom at the saddle-point.
Later in a one-dimensional diffusion model, Kramers \cite{Kramers40} derived a formula 
similar to this with a 
different factor before the integral. For large viscosity, the decay width is
\begin{gather}
\Gamma_{Kramers}(E^*,J) = f_k \,\, \Gamma_f^{BW}(E^*,J), \, \\ 
f_k = \sqrt{1+\left(\frac{\gamma}{2\omega}\right)^{2}} - \frac{\gamma}{\omega} , 
\end{gather}
where $\gamma$ is the magnitude of the viscosity and $\omega$ is the curvature 
of the potential energy at the saddle-point. The Kramers factor $f_k$ scaling the 
Bohr-Wheeler width is less than unity and is hard to extract experimentally 
due to the much larger uncertainty associated with the fission 
barrier and the level-density parameter. 

The fission decay width has also been suggested to be transient \cite{Grange83}, i.e. 
initially zero and then rising to the quasi-stationary value of Kramers. This
idea has helped to explain the larger number of neutrons emitted before the 
scission point is attained \cite{Hilscher92}.
During the transient time which can also be thought as a fission delay, 
any light-particle evaporation will lower the 
excitation energy and spin of the decaying nucleus and subsequently 
may reduce its fission probability.

However, there is some controversy as to whether  
transient fission decay widths are needed to explain experimental fission probabilities.
A number of theoretical  studies reproduce experimental fission probabilities 
and pre-scission neutron multiplicities with transient fission widths 
\cite{Frobrich93,Gontchar09}. The viscosity which determined the 
transient time scale was found to increase with the mass in these studies.
Transient fission has also been invoked to explain the unexpectedly
large number of evaporation residues measured in the very fissile
$^{216}$Th compound system formed in $^{32}$S+$^{184}$W reactions \cite{Back99}. 
Alternatively other studies have reproduced fission probabilities \cite{Moretto95} 
and both prescission neutron 
multiplicities and fission probabilities with no transient effects \cite{Lestone09}.
Similarly, in very-high-excitation-energy data obtained with 2.5-GeV proton induced spallation 
reactions, no transients were needed in reproducing the measured fission yields \cite{Tishchenko05}.

\begin{figure}[tbp]
\includegraphics*[scale=0.45]{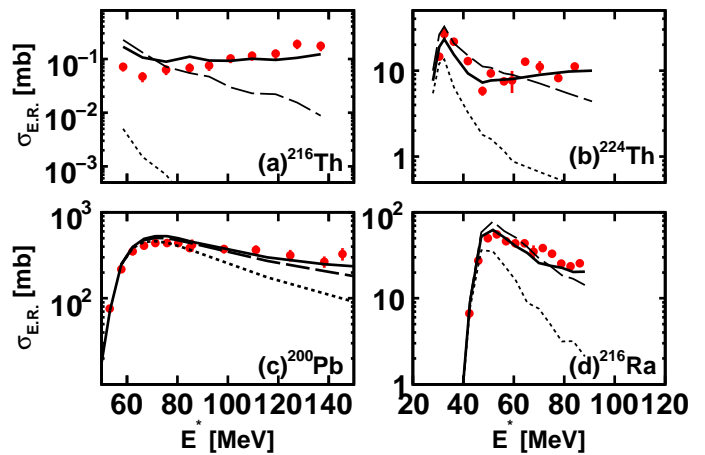}
\caption{(Color online) Evaporation-residue excitation functions for the
indicated compound nuclei. The data points are published experimental results
and the short-dashed, long-dashed, and solid curves were calculated with $\widetilde{a}_{eff}$=$A$/7.3, $\widetilde{a}$=$A$/11~MeV$^{-1}$ and 
Eq.~(\protect\ref{eq:fita}), respectively.}
\label{fig:fis}
\end{figure}

In this work, we will not try and answer all these uncertainties 
pertaining to fission, but will investigate 
how the excitation-dependent level-density parameter affects the fission probability.
The fission decay width will be taken from the Bohr-Wheeler formalism.
Let us  assume that the
level-density parameter for the saddle-point and ground-state configurations
are identical apart from a  scaling factor 
$a_{f}/a_{n}$ which accounts for the increased surface area of the former \cite{Toke81}.
Fission decay widths were calculated using the angular-momentum-dependent
fission barriers of Sierk \cite{Sierk86}. For $^{200}$Pb, $^{216}$Th,$^{224}$%
Th, and $^{224}$Ra compound nuclei formed in the reactions listed in Table~%
\ref{Tbl:fission}, both ER and fission excitation functions have been
measured allowing us to determined the fusion cross section and thus
constrain the CN spin distributions. Evaporation-residue
excitation functions were calculated with the exponential dependence of 
$\kappa$ in Fig.~\ref{fig:kappa} and some final adjustment was made with 
the parameter $a_{f}/a_{n}$ in order to reproduce the experimental data. 
The results, shown
by the solid curves in Fig.~\ref{fig:fis}, reproduce the data quite well and
the fitted $a_{f}/a_{n}$ values, which are all similar in magnitude,
 are listed in Table~\ref{Tbl:fission}. For
comparison, the short-dashed curves show the results obtained with a constant $%
\widetilde{a}_{eff}$=$A$/7.3~MeV$^{-1}$. The $U$ dependence of 
$\widetilde{a}_{eff}$ gives rise to an enhancement of the predicted ER yield
which is most pronounced for the heavier systems and the higher excitation
energies. However, for the energy regime where there is significant
enhancement, the fission cross sections are  orders of magnitude larger and
even with this enhancement, ER survival is still a rare process. 

\begin{table}[tbp]
\caption{Experimental data used in Fig~.\ref{fig:fis} are listed with
the compound nucleus, reaction, references and $a_f/a_n$ values used
in the {\sc GEMINI++} calculations.}
\label{Tbl:fission}%
\begin{ruledtabular}
\begin{tabular}{cD{+}{+}{4}ld}
 CN &     \multicolumn{1}{c}{reaction}  &  ref. & \multicolumn{1}{c}{$a_f/a_n$}\\
    &                                   &       &           \\
\hline
$^{200}$Pb&  ^{19}\mbox{F}+^{181}\mbox{Ta}& \cite{Hinde82,Caraley00,Fabris94}&1.04\\
$^{216}$Ra& ^{19}\mbox{F}+^{197}\mbox{Au}&\cite{Berriman01}&1.04\\
$^{216}$Th& ^{32}\mbox{S}+^{184}\mbox{W}&\cite{Back99,Keller87}&1.06\\
$^{224}$Th& ^{16}\mbox{O}+^{208}\mbox{Pb}& \cite{Videbaek77,Fineman94,Brinkmann94,Back85}&1.035\\
\end{tabular}
\end{ruledtabular}
\end{table}

The calculations with the excitation-dependent values of $\widetilde{a}_{eff}$ have higher 
nuclear temperatures than the $A$/7.3 MeV$^{-1}$ calculation. Larger temperatures enhance 
rare decay modes and these rare decay modes are the evaporation channels in these 
very fissile nuclei. This is illustrated by the long-dashed curves which are calculations 
with a constant $\tilde{a}_{eff}$=$A$/11~MeV$^{-1}$ where the temperatures are 
20\% larger than for $\widetilde{a}_{eff}$=$A$/7.3~MeV$^{-1}$. These curves also show
enhanced evaporation residue yields, but the excitation-energy dependence is
not as well described  as by the solid curves with the excitation-energy dependence. 
For the $^{216}$Th system of 
Fig.~\ref{fig:fis}(a),
Back \textit{et al.} using calculations with constant $\widetilde{a}_{eff}$ 
concluded 
that the statistical model was not able to reproduced the data and thus 
deduced that there must be  fission transients \cite{Back85}. However, it is now 
clear that with 
an excitation-energy dependent $\widetilde{a}_{eff}$, this conclusion is no longer valid.
This suggests possibility a  reduced role for fission transients in determining the fission
probability.

It should be noted that the ability of these calculations 
to reproduce the evaporation-residue cross sections depends on the assumed 
excitation-energy dependence of $\widetilde{a}_{eff}$. For the lighter $^{160}$Yb
system of Fig.~\ref{fig:yb160}, the excitation-energy dependence is rather 
well established \cite{Charity03}. A larger range of compound-nucleus excitation 
energies were probed (Table~\ref{Tbl:CN}) and neutron evaporation spectra were also measured.
Charged particles are typically emitted early the decay chain and probe 
higher excitation energies whereas neutrons are emitted at all decay stages and 
give information more on the average temperature. Reproduction of both charged-particle 
and neutron spectra required an excitation-energy dependence of 
$\widetilde{a}_{eff}$ for $^{160}$Yb.
Subsequently these $\widetilde{a}_{eff}$ values were found consistent with data from 
the similar-mass $^{178}$Hf compound nuclei at even lower excitation energies \cite{Komarov07}.
By contrast only charge-particle spectra were measured for the $^{224}$Th compound nucleus
in Ref.~\cite{Fineman94} and at just two excitation energies separated by $\sim$20~MeV.
It also was possible to fit these spectra with a constant $\widetilde{a}_{eff}$=$A$/15~MeV \cite{Fineman94}. 
Although a constant value is unlikely given the larger values derived from counting neutron resonances,
it is clear that for this heavy nucleus, the excitation-dependence of $\widetilde{a}_{eff}$ is not well 
constrained from the present experimental data. Clearly further experimental studies of this
point would be useful in understanding the fission in these very heavier systems.
Also it should be noted that for $A\sim$220, quasifission also competes with fusion 
reactions at the lower $\ell$ waves associated with evaporation-residue production for 
entrance channels with $^{19}$F projectiles and heavier \cite{Berriman01,Hinde02}.
This suggests that somewhat smaller values of $a_f/a_n$ are associated with 
the $^{19}$F+$^{197}$Au and $^{32}$S+$^{184}$W reactions than those of 
Table~\ref{Tbl:fission}.

Finally it is of interest to consider the relevance of this work to the
production of superheavy elements. Of particular interest are
``hot'' fusion reactions which have
produced the heaviest elements to date \cite{Oganessian04,Oganessian06}.
Based on an extrapolation of $\kappa $ to the $A=$277-294 region we would
 expect
significantly enhanced temperatures for the CN excitation energies of $\sim $%
35~MeV produced in these reactions. Therefore, this effect may also
contribute to an enhanced yield of superheavy elements in these hot fusion
reactions. Clearly, more studies are also needed in this area.

\subsection{Thermal Properties of Nuclei}
\label{Sec:thermal}

\begin{figure}[tbp]
\includegraphics*[scale=0.45]{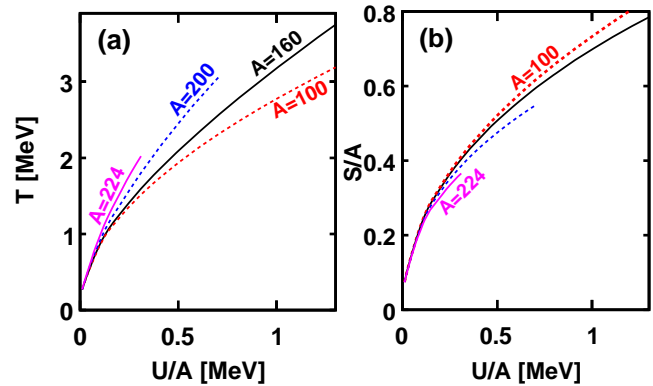}
\caption{(Color online) Excitation-energy dependence of (a) the nuclear
temperature and (b) the entropy deduced in this work.}
\label{fig:caloric}
\end{figure}

The thermal properties of nuclei can be inferred from the level density \cite%
{Bohr75}. Figures~\ref{fig:caloric}(a) and \ref%
{fig:caloric}(b) show the excitation-energy dependence of $S$ and $T$
plotted in a manner that the mass dependence would disappear for an energy-independent 
$\widetilde{a}_{eff}\propto A$. The curves for different masses are only plotted up to the
maximum $U$ sampled in the experiments. We see a small mass dependence of 
$S/A$, but a larger dependence for the temperature. For a given $U/A$, we see
smaller values of $S/A$ and larger temperatures for the 
heavier systems. The larger
temperatures are responsible for the stiffer evaporation spectra and the
enhancements of the small ER survival probabilities. 

The theoretical understanding of the rapid increase in $\kappa $ with $A$ is
not clear. Shlomo and Natowitz \cite{Shlomo91} assumed the effects of
long-range correlations wash out when $T$ becomes similar
in magnitude to the collective energy $\hbar \omega _{i}$ of each of the
modes. For many collective modes, $\omega _{i}$ varies
approximately inversely with the linear dimension, 
i.e. $\omega _{i}\sim A^{-1/3}$.
Values of $\kappa$ extracted from the predictions of Shlomo and Natowitz 
\cite{Shlomo91}, shown by 
the dashed curve in Fig.~\ref{fig:kappa}, have only a gentle mass dependence
and do not reproduce our experimental points.

\section{LIGHTER NUCLEI AND YRAST ENERGIES}
\label{Sec:light}
Due to the exponential-like dependence of $\kappa$ on mass, it seems that 
the kinetic-energy spectra should described by an excitation-independent
level-density parameter $\widetilde{a}_{eff}$ for the lighter nuclei. However, 
light nuclei have their own complications 
as the spin dependence of $E_{yrast}$ can be quite strong. 
This can cause quite pronounced effects on the predicted 
spectra of $\alpha$ particles which can remove appreciable angular momentum
from the decaying system. Such effects can in principle be isolated 
if both proton or neutron spectra are also measured as nucleons tend 
to remove very little angular momentum and thus are much less sensitive to 
$E_{yrast}$. However for the lightest nuclei, 
there is a lot more data available for 
$\alpha$ particles than protons. 

\begin{figure}[tbp]
\includegraphics*[scale=0.4]{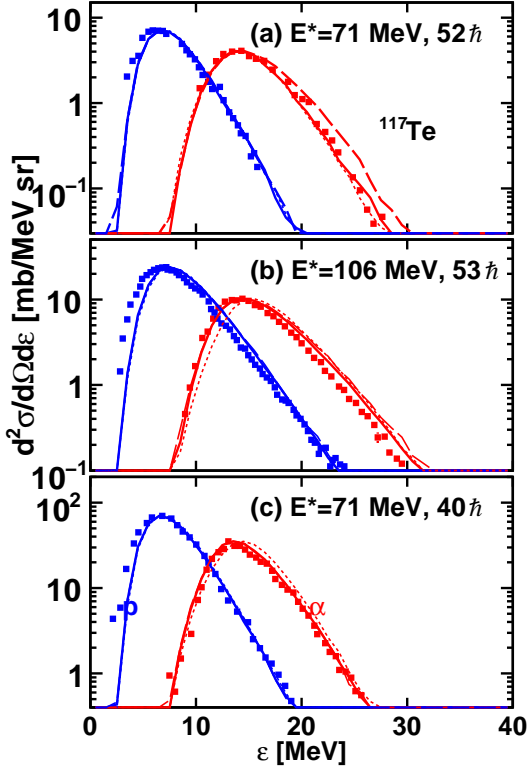}
\caption{(Color online) Inclusive proton and $\alpha$-particle kinetic spectra 
in the reaction center-of-mass frame measured at angles that highlight
compound-nucleus emission. The data are associated with $^{117}$Te compound  
nucleus formed in (a,b) $^{14}$N + $^{103}$Rh and (c) $^{40}$Ar+$^{77}$Se 
reactions. The curves are again {\sc GEMINI++} predictions. The solid curves 
are the default calculations with a distribution of Coulomb barriers,  
the excitation-dependent level-density parameter $\widetilde{a}_{eff}$, 
and the 
prescription for $E_{yrast}(J)$. For the short-dashed curves a single 
Coulomb barrier is used and for the long-dashed curves, Sierk's values of 
$E_{yarst}(J)$ are employed.}   
\label{fig:te117}
\end{figure}

\begin{figure}[tbp]
\includegraphics*[scale=0.4]{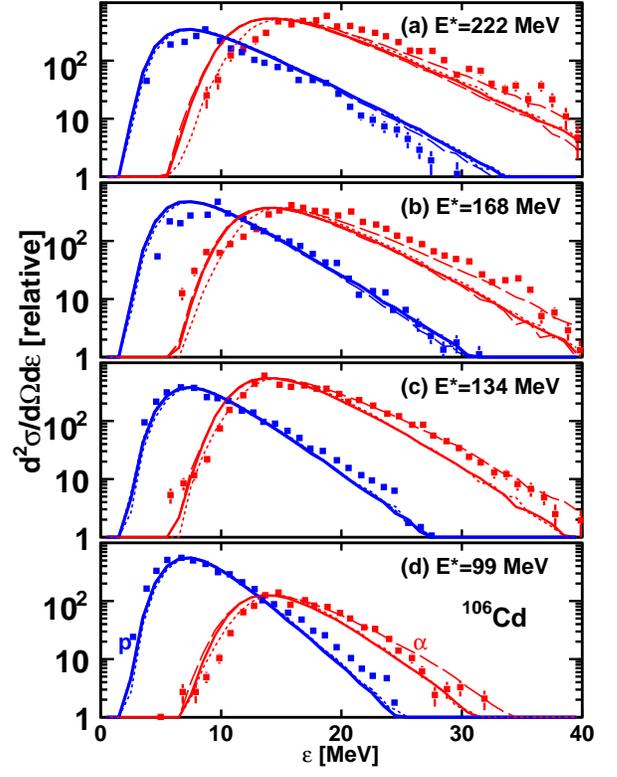}
\caption{(Color online) As for Fig.~\ref{fig:te117} but for $^{106}$Cd 
compound nuclei formed in $^{32}$S+$^{74}$Ge reactions. }
\label{fig:cd106}
\end{figure}

\begin{figure}[tbp]
\includegraphics*[scale=0.4]{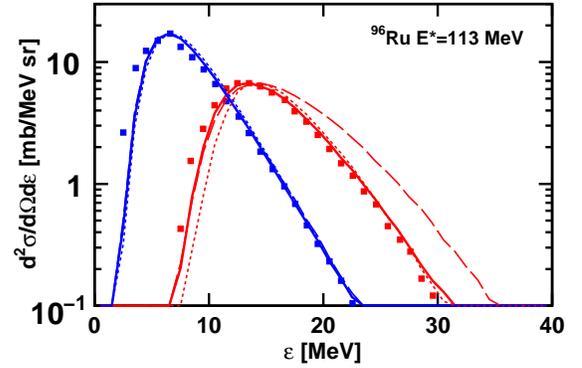}
\caption{(Color online) As for Fig.~\ref{fig:te117} but for $^{96}$Ru 
compound nuclei formed in $^{32}$S+$^{64}$Ni reactions. }
\label{fig:ru96}
\end{figure}

\begin{figure}[tbp]
\includegraphics*[scale=0.4]{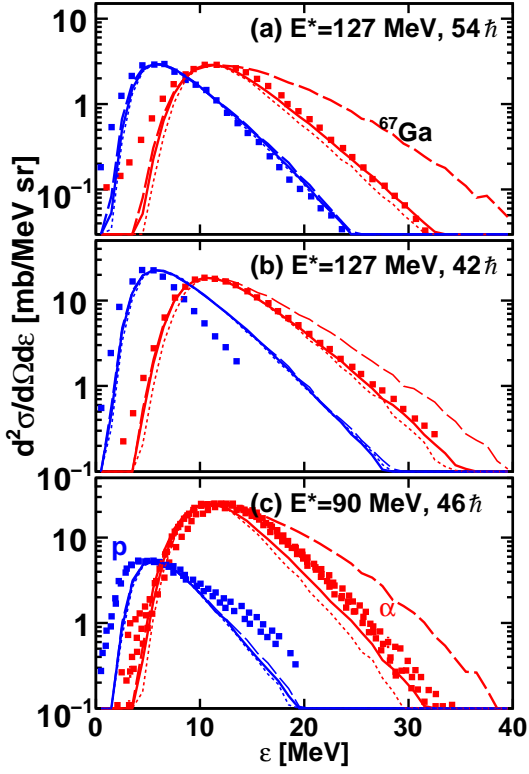}
\caption{(Color online) As for Fig.~\ref{fig:te117} but for $^{67}$Ga 
compound nuclei formed in $^{40}$Ar+$^{27}$Al and $^{55}$Mn+$^{12}$C
 reactions. }
\label{fig:ga67}
\end{figure}

\begin{figure}[tbp]
\includegraphics*[scale=0.4]{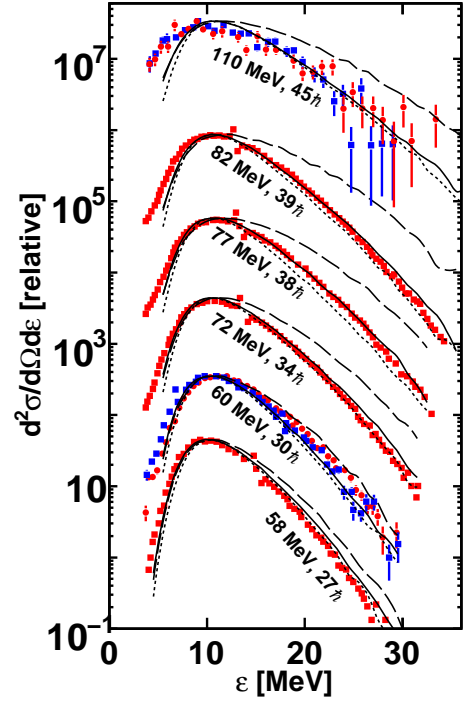}
\caption{(Color online) As for Fig.~\ref{fig:ga67}, but now for 
$^{59}$Cu compound nuclei.
For the $E^*$=60-MeV data, the square and circular data points represent 
the results measured at $\theta_{lab}$=25$^{\circ}$ and 45$^{\circ}$,
respectively, while for the 110-MeV data they correspond to $\theta_{lab}$=15$^{\circ}$ 
and 30$^{\circ}$.}    
 \label{fig:cu59}
\end{figure}

Let us concentrate on the spectra
for $A$=117 to 59 compound nuclei in Figs.~\ref{fig:te117} to \ref{fig:cu59}.
{\sc GEMINI++} calculations including the distribution of Coulomb barrier,
Sierk's values of $E_{yarst}$, and 
the excitation-energy-dependent level-density parameter are indicated by the 
long-dashed curves. Calculations with a constant $\widetilde{a}_{eff}=A/7.3$~MeV$^{-1}$ 
would be essentially identical to these. For protons with minimal 
angular-momentum effects, one obtained good agreement with experimental data 
for the $^{117}$Te, 
$^{106}$Cd, and $^{96}$Ru compound systems in Figs.~\ref{fig:te117} to 
\ref{fig:ru96}. For the $^{67}$Ga system, the proton spectra are not very well
reproduced in Figs.~\ref{fig:ga67}(b) and \ref{fig:ga67}(c). Actually is 
difficult to understand the evolution of the slope of the exponential 
tails of these proton spectra with excitation energy within the statistical 
model. Possibility there are experimental problems here or there is 
contamination from other processes. In fact for all these lighter nuclei 
the possibility of contamination exists as the data are all inclusive. 
 
Consider the $^{59}$Cu data from the  
$^{32}$S+$^{27}$Al reaction if Fig.~\ref{fig:cu59}. The evaporation-residue
cross section represents about 85\% of the total reaction cross section at 
$E_{beam}$=100~MeV ($E^*$=58~MeV)
but decreases to 46\% at $E_{beam}$=214~MeV ($E^*$=110~MeV) \cite{Doukellis88}. 
The remaining component of the reaction cross
section is associated with binary-reaction dynamics with various degrees of damping and
these binary-reaction products evaporate protons and $\alpha$ particles 
\cite{Winkler81,Pelte81,Manduchi85}. Very damped binary 
and fusion-fission reactions are associated with extensive angular distributions and 
thus light-particle emission from these processes will not have a strong angular 
distribution and would be difficult to separate from those associated with evaporation residues.
Clearly not all the inclusive $\alpha$ and \textit{p}
spectra can be associated with evaporation as is assumed in most analyses. The exact 
extent of this contamination from binary reactions has not been established, but in this work,
it will be assumed that it is not large for $\alpha$ particles and the basic features of the 
spectra can be traced to 
evaporation from the fused system. 

For $\alpha$ particles, the {\sc GEMINI++} predictions significantly over estimate the yield 
in the high-energy tail for many of the data sets. In fact these predicted spectra
do not have exponential tails in the sense that the spectral tails  decrease linearly on a 
log plot. This is an indication that the predicted enhancement of the high-energy region 
is not a consequence of high temperatures, but of angular-momentum effects associated the 
steep increase of $E_{yrast}$ with $J$. The angular-momentum effects are most pronounced
for the more symmetric reactions such as the $^{40}$Ar+$^{27}$Al  
reactions in Figs.~\ref{fig:ga67}(a) and \ref{fig:ga67}(c) which populate a region of 
$E^*-J$ space near the yrast line at high spins. One also finds the same for
 the higher-energy $^{32}$S+$^{27}$Al reactions in Fig.~\ref{fig:cu59}.    

A large number of previous studies have noted
that calculations with Sierk's or the RLDM values of $E_{yarst}$ are incapable of 
reproducing $\alpha$-particle spectra from light systems with large angular momentum
 \cite{LaRana87,Choudhury84,Majka87,Fornal88,Viesti88,Huizenga89,Fornal90,Kildir92,Brown99}.

Huizenga \textit{et al.} \cite{Huizenga89} reproduced  experimental $\alpha$-particle spectra
by using a modified yrast energy given by   
\begin{equation}
E_{yrast}(J) = \frac{\hbar^2}{2 \mathcal{I}_{rig}}(1+\delta_1 J^2 + \delta_2 J^4)
\label{eq:Huizenga}
\end{equation}
which contains two free parameters, $\delta_2$ and $\delta_4$ adjusted for each compound nucleus.
Equally good fits to the data can be obtained by using the Sierk calculations out to an
angular momentum $J_*$ and subsequently allowing $E_{yrast}(J)$ to increase linearly for higher spins, i.e.,
\begin{equation}
E_{yrast}(J) =
\begin{cases}
E_{Sierk}(J) & \text{if } J < J_{*} \\
E_{Sierk}(J_{*}) + (J-J_{*}) E_{Sierk}'(J_{*})& \text{if } J > J_{*} \\
\end{cases}
\label{Eq:modYrast}
\end{equation}
This has the advantage of having only one free parameter making interpolation and 
extrapolation easier. Also with increasingly large values of $J_*$, the effect turns off as Sierk's
calculations become more linear (see later). In addition if $J_*$ is made larger than the input 
compound-nucleus spin distribution it has not effect. Thus if $J_*$ increases with $A$, it allows a 
smooth transition to heavier nucleus where Sierk's values can reproduce experimental data.

\begin{figure}[tbp]
\includegraphics*[scale=0.45]{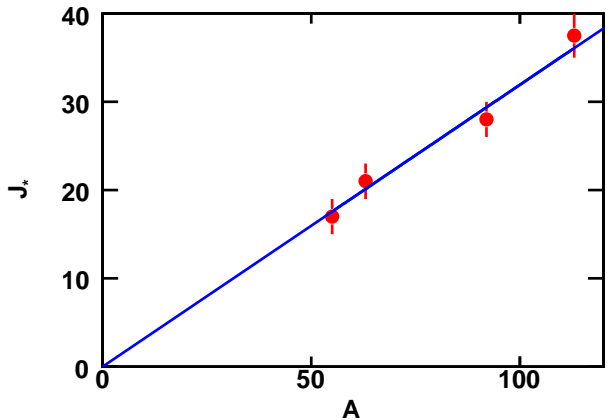}
\caption{(Color online) Values of $J_*$, the angular momentum for which the 
Sierk yrast energy is modified, are plotted again the mass of the 
first $\alpha$-daughter nucleus. The line shows a fits to these values
which is used in subsequent {\sc GEMINI++} calculations.}
\label{fig:jstar}
\end{figure}

The value of $J_*$ was obtained from fits to the data 
from  $^{59}$Cu, $^{67}$Ga, $^{96}$Ru, and $^{117}$Te  compound nuclei and 
the values are plotted against the  $A$ of the $\alpha$-daughter system in Fig.~\ref{fig:jstar}.
These data points can be fit with the linear function 
\begin{equation}
J_{*} = 0.319 A
\end{equation}
shown by the solid line. {\sc GEMINI++} predictions with 
this global parametrization of 
$J_*$ are shown by the solid curves in Figs.~\ref{fig:te117} 
to \ref{fig:cu59} and reproduce
the experimental distributions reasonably well. The exception is for the 
$^{106}$Cd compound nucleus where the original long-dashed calculations
in Fig.~\ref{fig:cd106} 
obtained with Sierk's $E_{yrast}(J)$ values produced a better fit.

\begin{figure}[tbp]
\includegraphics*[scale=0.45]{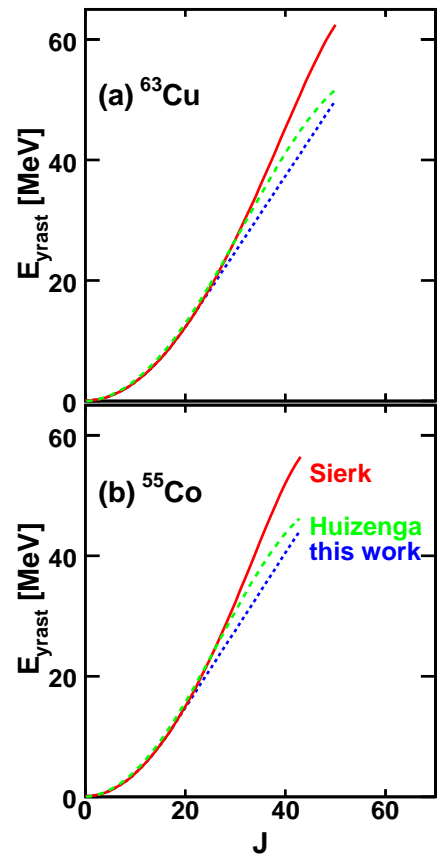}
\caption{(Color online) Rotational-plus-deformation energies verses the 
nuclear angular momentum for (a) $^{63}$Cu and (b) $^{55}$Co. Curves are shown 
for the dependence caluculated by Sierk form his macroscopioc model. 
These can be compared to results obtain from fitting $\alpha$-particle evaporation spectra in this work and by Huizenga \textit{et al.}}  
\label{fig:yrast}
\end{figure}

In Fig.~\ref{fig:yrast}, we compare the modified $E_{yrast}$ energies to Sierk's
calculations for $^{63}$Cu and $^{55}$Co, the daughter nuclei following $\alpha$ 
evaporation from the $^{67}$Ga and $^{59}$Cu compound nuclei. In addition are shown
values obtained by Huizenga \textit{et al.} obtained from fitting these data with 
Eq.~(\ref{eq:Huizenga}) \cite{Huizenga89}. 
Although the values from this work are slightly lower than those of Huizenga \textit{et al.} at 
the high spins, the most important comparison is that the slopes of $E_{yrast}(J)$
are very similar at these high spins. As mentioned before, evaporation spectra are 
not sensitive to absolute level density. In this case the calculations are sensitive 
to the $J$ dependence of $\rho$ which is dictated by the spin dependence of $E_{yrast}$.
The $E_{yrast}$ values of Huizenga would also give good reproduction of the experimental
data if they were used in {\sc GEMINI++}. Evaporation spectra thus give information on the 
$J$ dependence of $E_{yrast}$.

Huizenga \textit{et al} also suggested that $E_{yrast}$ at these high spins not be interpreted as
just the rotational-plus-deformation energy of the nucleus after shell and pairing effects have vanished.
Rather they should be treated as effective values that may take account of other effects such as
a spin dependence of the level-density parameter, or spin dependence of collective enhancements.

In Figs.~\ref{fig:te117} to \ref{fig:cu59}, the short-dashed curves again show the 
predictions with no barrier distributions in the transmission coefficients. 
As the absolute barriers are 
smaller for these lighter nuclei, the effect of the barrier distributions
on the spectra are reduced. 
However, the inclusion of the distributions 
(solid curves) still improves the agreement with the $\alpha$-particle 
data except for the $^{106}$Cd compound nucleus where the long-dashed curves 
give better fits. The $^{106}$Cd data has thus proven exceptional in the 
ingredients necessary to fit both the exponential tails and the Coulomb barrier 
region. The standard calculations represented by the solid curves in 
Fig.~\ref{fig:cd106} would fit much better if the experimental spectra were 
shifted down in energy.

For the $^{59}$Cu and $^{67}$Ga systems, it is clear that the enhancement from the barrier
distributions at the largest excitation energies and $J_0$ values 
is not sufficient to reproduce the experimental $\alpha$-particle spectra 
in Fig.~\ref{fig:cu59}. Again there are questions about contamination
from events not associated with evaporation residues.
Majka \textit{et al.} \cite{Majka87} have investigated the need for
a $J$ dependence of the transmission coefficients which they associate with 
the increasing deformation of the equilibrium configuration with spin.
At present we have not attempted to modify the transmission coefficients
as a function of $J$ in {\sc GEMINI++} to better reproduce the data.

For the proton spectra, the inclusion of the barrier distribution 
practically has no effect. However, the subbarrier region in the 
$^{117}$Te (Fig.~\ref{fig:te117}), $^{96}$Ru (Fig.~\ref{fig:ru96})
are still underestimated in the calculations even though 
the rest of the spectral
shape is well described. Again, there are questions as to whether this is 
a problem with contamination from other processes. 
Alternatively, enhancements to the proton subbarrier region 
can also arise if the evaporation residue is sufficiently proton rich. 
If  the decay chain of particles leads to a daughter nucleus with excitation 
energy below the neutron separation energy, but above the proton value, then
subbarrier proton emission competes with $\gamma$ emission. Such protons are 
the source of the lowest-energy protons in the {\sc GEMINI++} predictions 
for these systems.

\section{CONCLUSION}

A systematic review of the ingredient necessary to describe the shape of 
proton and $\alpha$-particle, and some neutron  evaporation spectra was made.
In order to describe the low-energy yields of the charged particles, 
transmission coefficients associated with a distribution of barriers were 
necessary. This was incorporated in a simple way into the statistical model
assuming a distribution of barriers which was assumed to arise 
from large thermal fluctuations.
This could include fluctuations in shape, density, or surface diffuseness.

The nuclear level density was described in terms of the Fermi-gas formula
which is valid for single-particle excitations. However an effective 
level-density parameter is used that can also account for collective contributions.
For light nucleus ($A<$120), the shell-smoothed values of 
$\widetilde{a}$=$A$/7.3~MeV$^{-1}$,
obtained from neutron-resonance counting at low excitation energies, 
was also found consistent with the evaporation spectra. However for heavier
nuclei at large excitation energies, smaller level-density parameters are needed. 
Evaporation spectra
were fit with an excitation-energy-dependent level-density parameter
where the excitation-energy dependence increases very rapidly with $A$.
This excitation-energy dependence was also found important in understanding
the survival against fission in very fissile nuclei and allowed reproduction 
of data that previously was thought to require fission transients.

The angular-momentum dependence of the level-density is largely defined by 
the spin-dependence of the macroscopic yrast energy. 
For light compound nuclei at large $J$, modifications to Sierk's and the 
rotating-liquid-drop model values of the rotation-plus-deformation energies
which reduces the angular-momentum dependence of the level density 
were needed to describe experimental $\alpha$-particle evaporation spectra.
 
These ingredients were incorporated in the {\sc GEMINI++} code to allow
a good description of the spectral shape of evaporation spectra over all of 
the periodic table.

This work was supported by the U.S. Department of Energy, Division of
Nuclear Physics under grant DE-FG02-87ER-40316.


\begin{thebibliography}{10}%
\makeatletter
\providecommand \@ifxundefined [1]{%
 \ifx #1\undefined \expandafter \@firstoftwo
 \else \expandafter \@secondoftwo
\fi
}%
\providecommand \@ifnum [1]{%
 \ifnum #1\expandafter \@firstoftwo
 \else \expandafter \@secondoftwo
\fi
}%
\providecommand \enquote [1]{``#1''}%
\providecommand \bibnamefont  [1]{#1}%
\providecommand \bibfnamefont [1]{#1}%
\providecommand \citenamefont [1]{#1}%
\providecommand\href[0]{\@sanitize\@href}%
\providecommand\@href[1]{\endgroup\@@startlink{#1}\endgroup\@@href}%
\providecommand\@@href[1]{#1\@@endlink}%
\providecommand \@sanitize [0]{\begingroup\catcode`\&12\catcode`\#12\relax}%
\@ifxundefined \pdfoutput {\@firstoftwo}{%
 \@ifnum{\z@=\pdfoutput}{\@firstoftwo}{\@secondoftwo}%
}{%
 \providecommand\@@startlink[1]{\leavevmode}%
 \providecommand\@@endlink[0]{}%
}{%
 \providecommand\@@startlink[1]{%
  \leavevmode
  \pdfstartlink
   attr{/Border[0 0 1 ]/H/I/C[0 1 1]}%
   user{/Subtype/Link/A<</Type/Action/S/URI/URI(#1)>>}%
  \relax
 }%
 \providecommand\@@endlink[0]{\pdfendlink}%
}%
\providecommand \url  [0]{\begingroup\@sanitize \@url }%
\providecommand \@url [1]{\endgroup\@href {#1}{\urlprefix}}%
\providecommand \urlprefix [0]{URL }%
\providecommand \Eprint[0]{\href }%
\@ifxundefined \urlstyle {%
  \providecommand \doi [1]{doi:\discretionary{}{}{}#1}%
}{%
  \providecommand \doi [0]{doi:\discretionary{}{}{}\begingroup
  \urlstyle{rm}\Url }%
}%
\providecommand \doibase [0]{http://dx.doi.org/}%
\providecommand \Doi[1]{\href{\doibase#1}}%
\providecommand \bibAnnote [3]{%
  \BibitemShut{#1}%
  \begin{quotation}\noindent
    \textsc{Key:}\ #2\\\textsc{Annotation:}\ #3%
  \end{quotation}%
}%
\providecommand \bibAnnoteFile [2]{%
  \IfFileExists{#2}{\bibAnnote {#1} {#2} {\input{#2}}}{}%
}%
\providecommand \typeout [0]{\immediate \write \m@ne }%
\providecommand \selectlanguage [0]{\@gobble}%
\providecommand \bibinfo [0]{\@secondoftwo}%
\providecommand \bibfield [0]{\@secondoftwo}%
\providecommand \translation [1]{[#1]}%
\providecommand \BibitemOpen[0]{}%
\providecommand \bibitemStop [0]{}%
\providecommand \bibitemNoStop [0]{.\EOS\space}%
\providecommand \EOS [0]{\spacefactor3000\relax}%
\providecommand \BibitemShut [1]{\csname bibitem#1\endcsname}%
\bibitem{Yariv81}%
  \BibitemOpen
  \bibfield{author}{%
  \bibinfo {author} {\bibfnamefont{Y.}~\bibnamefont{Yariv}}\ and\ \bibinfo
  {author} {\bibfnamefont{Z.}~\bibnamefont{Fraenkel}},\ }%
  \bibfield{journal}{%
  \Doi{10.1103/PhysRevC.24.488}{\bibinfo {journal} {Phys. Rev. C}}\ }%
  \textbf{\bibinfo {volume} {24}},\ \bibinfo {pages} {488} (\bibinfo {year}
  {1981})%
  \bibAnnoteFile{NoStop}{Yariv81}%
\bibitem{Boudard02}%
  \BibitemOpen
  \bibfield{author}{%
  \bibinfo {author} {\bibfnamefont{A.}~\bibnamefont{Boudard}}, \bibinfo
  {author} {\bibfnamefont{J.}~\bibnamefont{Cugnon}}, \bibinfo {author}
  {\bibfnamefont{S.}~\bibnamefont{Leray}},\ and\ \bibinfo {author}
  {\bibfnamefont{C.}~\bibnamefont{Volant}},\ }%
  \bibfield{journal}{%
  \Doi{10.1103/PhysRevC.66.044615}{\bibinfo {journal} {Phys. Rev. C}}\ }%
  \textbf{\bibinfo {volume} {66}},\ \bibinfo {pages} {044615} (\bibinfo {year}
  {2002})%
  \bibAnnoteFile{NoStop}{Boudard02}%
\bibitem{ou09}%
  \BibitemOpen
  \bibfield{author}{%
  \bibinfo {author} {\bibfnamefont{L.}~\bibnamefont{Ou}}, \bibinfo {author}
  {\bibfnamefont{Z.}~\bibnamefont{Li}}, \bibinfo {author}
  {\bibfnamefont{X.}~\bibnamefont{Wu}}, \bibinfo {author}
  {\bibfnamefont{J.}~\bibnamefont{Tian}},\ and\ \bibinfo {author}
  {\bibfnamefont{W.}~\bibnamefont{Sun}},\ }%
  \bibfield{journal}{%
  \bibinfo {journal} {J. Phys. G: Nucl. Part. Phys.}\ }%
  \textbf{\bibinfo {volume} {36}},\ \bibinfo {pages} {125104} (\bibinfo {year}
  {2009})%
  \bibAnnoteFile{NoStop}{ou09}%
\bibitem{Cinausero96}%
  \BibitemOpen
  \bibfield{author}{%
  \bibinfo {author} {\bibfnamefont{M.}~\bibnamefont{Cinausero}}, \bibinfo
  {author} {\bibfnamefont{G.}~\bibnamefont{Prete}}, \bibinfo {author}
  {\bibfnamefont{D.}~\bibnamefont{Fabris}}, \bibinfo {author}
  {\bibfnamefont{G.}~\bibnamefont{Nebbia}}, \bibinfo {author}
  {\bibfnamefont{G.}~\bibnamefont{Viesti}}, \bibinfo {author}
  {\bibfnamefont{G.~X.}\ \bibnamefont{Dai}}, \bibinfo {author}
  {\bibfnamefont{K.}~\bibnamefont{Hagel}}, \bibinfo {author}
  {\bibfnamefont{J.}~\bibnamefont{Li}}, \bibinfo {author}
  {\bibfnamefont{Y.}~\bibnamefont{Lou}}, \bibinfo {author}
  {\bibfnamefont{J.~B.}\ \bibnamefont{Natowitz}}, \bibinfo {author}
  {\bibfnamefont{D.}~\bibnamefont{Utley}}, \bibinfo {author}
  {\bibfnamefont{R.}~\bibnamefont{Wada}}, \bibinfo {author}
  {\bibfnamefont{N.}~\bibnamefont{Gelli}}, \bibinfo {author}
  {\bibfnamefont{F.}~\bibnamefont{Lucarelli}},\ and\ \bibinfo {author}
  {\bibfnamefont{M.}~\bibnamefont{Colonna}},\ }%
  \bibfield{journal}{%
  \Doi{DOI: 10.1016/0370-2693(96)00793-9}{\bibinfo {journal} {Phys. Lett. B}}\
  }%
  \textbf{\bibinfo {volume} {383}},\ \bibinfo {pages} {372 } (\bibinfo {year}
  {1996})%
  \bibAnnoteFile{NoStop}{Cinausero96}%
\bibitem{Charity97}%
  \BibitemOpen
  \bibfield{author}{%
  \bibinfo {author} {\bibfnamefont{R.~J.}\ \bibnamefont{Charity}}, \bibinfo
  {author} {\bibfnamefont{M.}~\bibnamefont{Korolija}}, \bibinfo {author}
  {\bibfnamefont{D.~G.}\ \bibnamefont{Sarantites}},\ and\ \bibinfo {author}
  {\bibfnamefont{L.~G.}\ \bibnamefont{Sobotka}},\ }%
  \bibfield{journal}{%
  \Doi{10.1103/PhysRevC.56.873}{\bibinfo {journal} {Phys. Rev. C}}\ }%
  \textbf{\bibinfo {volume} {56}},\ \bibinfo {pages} {873} (\bibinfo {year}
  {1997})%
  \bibAnnoteFile{NoStop}{Charity97}%
\bibitem{Liang97}%
  \BibitemOpen
  \bibfield{author}{%
  \bibinfo {author} {\bibfnamefont{J.~F.}\ \bibnamefont{Liang}}, \bibinfo
  {author} {\bibfnamefont{J.~D.}\ \bibnamefont{Bierman}}, \bibinfo {author}
  {\bibfnamefont{M.~P.}\ \bibnamefont{Kelly}}, \bibinfo {author}
  {\bibfnamefont{A.~A.}\ \bibnamefont{Sonzogni}}, \bibinfo {author}
  {\bibfnamefont{R.}~\bibnamefont{Vandenbosch}},\ and\ \bibinfo {author}
  {\bibfnamefont{J.~P.~S.}\ \bibnamefont{van Schagen}},\ }%
  \bibfield{journal}{%
  \Doi{10.1103/PhysRevC.56.908}{\bibinfo {journal} {Phys. Rev. C}}\ }%
  \textbf{\bibinfo {volume} {56}},\ \bibinfo {pages} {908} (\bibinfo {year}
  {1997})%
  \bibAnnoteFile{NoStop}{Liang97}%
\bibitem{Charity08}%
  \BibitemOpen
  \bibfield{author}{%
  \bibinfo {author} {\bibfnamefont{R.~J.}\ \bibnamefont{Charity}},\ }%
  in\ \emph{\bibinfo {booktitle} {Joint {ICTP}-{AIEA} Advanced Workshop on
  Model Codes for Spallation Reactions}},\ \bibinfo {series and number}
  {\bibinfo {number} {Report INDC(NDC)-0530}}\ (\bibinfo {publisher} {IAEA},\
  \bibinfo {address} {Vienna},\ \bibinfo {year} {2008})%
  \bibAnnoteFile{NoStop}{Charity08}%
\bibitem{Charity88a}%
  \BibitemOpen
  \bibfield{author}{%
  \bibinfo {author} {\bibfnamefont{R.~J.}\ \bibnamefont{Charity}}, \bibinfo
  {author} {\bibfnamefont{M.~A.}\ \bibnamefont{McMahan}}, \bibinfo {author}
  {\bibfnamefont{G.~J.}\ \bibnamefont{Wozniak}}, \bibinfo {author}
  {\bibfnamefont{R.~J.}\ \bibnamefont{McDonald}}, \bibinfo {author}
  {\bibfnamefont{L.~G.}\ \bibnamefont{Moretto}}, \bibinfo {author}
  {\bibfnamefont{D.~G.}\ \bibnamefont{Sarantites}}, \bibinfo {author}
  {\bibfnamefont{L.~G.}\ \bibnamefont{Sobotka}}, \bibinfo {author}
  {\bibfnamefont{G.}~\bibnamefont{Guarino}}, \bibinfo {author}
  {\bibfnamefont{A.}~\bibnamefont{Panteleo}}, \bibinfo {author}
  {\bibfnamefont{L.}~\bibnamefont{Fiore}}, \bibinfo {author}
  {\bibfnamefont{A.}~\bibnamefont{Gobbi}},\ and\ \bibinfo {author}
  {\bibfnamefont{K.}~\bibnamefont{Hildenbrand}},\ }%
  \bibfield{journal}{%
  \bibinfo {journal} {Nucl. Phys.}\ }%
  \textbf{\bibinfo {volume} {A483}},\ \bibinfo {pages} {371} (\bibinfo {year}
  {1988})%
  \bibAnnoteFile{NoStop}{Charity88a}%
\bibitem{Komarov07}%
  \BibitemOpen
  \bibfield{author}{%
  \bibinfo {author} {\bibfnamefont{S.}~\bibnamefont{Komarov}}, \bibinfo
  {author} {\bibfnamefont{R.~J.}\ \bibnamefont{Charity}}, \bibinfo {author}
  {\bibfnamefont{C.~J.}\ \bibnamefont{Chiara}}, \bibinfo {author}
  {\bibfnamefont{W.}~\bibnamefont{Reviol}}, \bibinfo {author}
  {\bibfnamefont{D.~G.}\ \bibnamefont{Sarantites}}, \bibinfo {author}
  {\bibfnamefont{L.~G.}\ \bibnamefont{Sobotka}}, \bibinfo {author}
  {\bibfnamefont{A.~L.}\ \bibnamefont{Caraley}}, \bibinfo {author}
  {\bibfnamefont{M.~P.}\ \bibnamefont{Carpenter}},\ and\ \bibinfo {author}
  {\bibfnamefont{D.}~\bibnamefont{Seweryniak}},\ }%
  \bibfield{journal}{%
  \Doi{10.1103/PhysRevC.75.064611}{\bibinfo {journal} {Phys. Rev. C}}\ }%
  \textbf{\bibinfo {volume} {75}},\ \bibinfo {eid} {064611} (\bibinfo {year}
  {2007})%
  \bibAnnoteFile{NoStop}{Komarov07}%
\bibitem{Bass74}%
  \BibitemOpen
  \bibfield{author}{%
  \bibinfo {author} {\bibfnamefont{R.}~\bibnamefont{Bass}},\ }%
  \bibfield{journal}{%
  \Doi{DOI: 10.1016/0375-9474(74)90292-9}{\bibinfo {journal} {Nucl. Phys.}}\ }%
  \textbf{\bibinfo {volume} {A231}},\ \bibinfo {pages} {45 } (\bibinfo {year}
  {1974})%
  \bibAnnoteFile{NoStop}{Bass74}%
\bibitem{Bass77}%
  \BibitemOpen
  \bibfield{author}{%
  \bibinfo {author} {\bibfnamefont{R.}~\bibnamefont{Bass}},\ }%
  \bibfield{journal}{%
  \Doi{10.1103/PhysRevLett.39.265}{\bibinfo {journal} {Phys. Rev. Lett.}}\ }%
  \textbf{\bibinfo {volume} {39}},\ \bibinfo {pages} {265} (\bibinfo {year}
  {1977})%
  \bibAnnoteFile{NoStop}{Bass77}%
\bibitem{Mancusi10}%
  \BibitemOpen
  \bibfield{author}{%
  \bibinfo {author} {\bibfnamefont{D.}~\bibnamefont{Mancusi}}, \bibinfo
  {author} {\bibfnamefont{R.~J.}\ \bibnamefont{Charity}},\ and\ \bibinfo
  {author} {\bibfnamefont{J.}~\bibnamefont{Cugnon}},\ }%
  \enquote{\bibinfo {title} {Unified description of nuclear de-excitation in
  fusion and spallation reactions},}\  (\bibinfo {year} {2010}),\ \bibinfo
  {note} {to be published}%
  \bibAnnoteFile{NoStop}{Mancusi10}%
\bibitem{Fornal88}%
  \BibitemOpen
  \bibfield{author}{%
  \bibinfo {author} {\bibfnamefont{B.}~\bibnamefont{Fornal}}, \bibinfo {author}
  {\bibfnamefont{G.}~\bibnamefont{Prete}}, \bibinfo {author}
  {\bibfnamefont{G.}~\bibnamefont{Nebbia}}, \bibinfo {author}
  {\bibfnamefont{F.}~\bibnamefont{Trotti}}, \bibinfo {author}
  {\bibfnamefont{G.}~\bibnamefont{Viesti}}, \bibinfo {author}
  {\bibfnamefont{D.}~\bibnamefont{Fabris}}, \bibinfo {author}
  {\bibfnamefont{K.}~\bibnamefont{Hagel}},\ and\ \bibinfo {author}
  {\bibfnamefont{J.~B.}\ \bibnamefont{Natowitz}},\ }%
  \bibfield{journal}{%
  \Doi{10.1103/PhysRevC.37.2624}{\bibinfo {journal} {Phys. Rev. C}}\ }%
  \textbf{\bibinfo {volume} {37}},\ \bibinfo {pages} {2624} (\bibinfo {year}
  {1988})%
  \bibAnnoteFile{NoStop}{Fornal88}%
\bibitem{Gutbrod73}%
  \BibitemOpen
  \bibfield{author}{%
  \bibinfo {author} {\bibfnamefont{H.~H.}\ \bibnamefont{Gutbrod}}, \bibinfo
  {author} {\bibfnamefont{W.~G.}\ \bibnamefont{Winn}},\ and\ \bibinfo {author}
  {\bibfnamefont{M.}~\bibnamefont{Blann}},\ }%
  \bibfield{journal}{%
  \Doi{DOI: 10.1016/0375-9474(73)90149-8}{\bibinfo {journal} {Nucl. Phys.}}\ }%
  \textbf{\bibinfo {volume} {A213}},\ \bibinfo {pages} {267 } (\bibinfo {year}
  {1973})%
  \bibAnnoteFile{NoStop}{Gutbrod73}%
\bibitem{Kozub75}%
  \BibitemOpen
  \bibfield{author}{%
  \bibinfo {author} {\bibfnamefont{R.~L.}\ \bibnamefont{Kozub}}, \bibinfo
  {author} {\bibfnamefont{N.~H.}\ \bibnamefont{Lu}}, \bibinfo {author}
  {\bibfnamefont{J.~M.}\ \bibnamefont{Miller}}, \bibinfo {author}
  {\bibfnamefont{D.}~\bibnamefont{Logan}}, \bibinfo {author}
  {\bibfnamefont{T.~W.}\ \bibnamefont{Debiak}},\ and\ \bibinfo {author}
  {\bibfnamefont{L.}~\bibnamefont{Kowalski}},\ }%
  \bibfield{journal}{%
  \Doi{10.1103/PhysRevC.11.1497}{\bibinfo {journal} {Phys. Rev. C}}\ }%
  \textbf{\bibinfo {volume} {11}},\ \bibinfo {pages} {1497} (\bibinfo {year}
  {1975})%
  \bibAnnoteFile{NoStop}{Kozub75}%
\bibitem{Puhlhofer77}%
  \BibitemOpen
  \bibfield{author}{%
  \bibinfo {author} {\bibfnamefont{F.}~\bibnamefont{P{\"u}hlhofer}}, \bibinfo
  {author} {\bibfnamefont{W.~F.~W.}\ \bibnamefont{Schneider}}, \bibinfo
  {author} {\bibfnamefont{F.}~\bibnamefont{Busch}}, \bibinfo {author}
  {\bibfnamefont{J.}~\bibnamefont{Barrette}}, \bibinfo {author}
  {\bibfnamefont{P.}~\bibnamefont{Braun-Munzinger}}, \bibinfo {author}
  {\bibfnamefont{C.~K.}\ \bibnamefont{Gelbke}},\ and\ \bibinfo {author}
  {\bibfnamefont{H.~E.}\ \bibnamefont{Wegner}},\ }%
  \bibfield{journal}{%
  \Doi{10.1103/PhysRevC.16.1010}{\bibinfo {journal} {Phys. Rev. C}}\ }%
  \textbf{\bibinfo {volume} {16}},\ \bibinfo {pages} {1010} (\bibinfo {year}
  {1977})%
  \bibAnnoteFile{NoStop}{Puhlhofer77}%
\bibitem{Rosner85}%
  \BibitemOpen
  \bibfield{author}{%
  \bibinfo {author} {\bibfnamefont{G.}~\bibnamefont{Rosner}}, \bibinfo {author}
  {\bibfnamefont{J.}~\bibnamefont{Pochodzalla}}, \bibinfo {author}
  {\bibfnamefont{B.}~\bibnamefont{Heck}}, \bibinfo {author}
  {\bibfnamefont{G.}~\bibnamefont{Hlawatsch}}, \bibinfo {author}
  {\bibfnamefont{A.}~\bibnamefont{Miczaika}}, \bibinfo {author}
  {\bibfnamefont{H.~J.}\ \bibnamefont{Rabe}}, \bibinfo {author}
  {\bibfnamefont{R.}~\bibnamefont{Butsch}}, \bibinfo {author}
  {\bibfnamefont{B.}~\bibnamefont{Kolb}},\ and\ \bibinfo {author}
  {\bibfnamefont{B.}~\bibnamefont{Sedelmeyer}},\ }%
  \bibfield{journal}{%
  \Doi{DOI: 10.1016/0370-2693(85)90144-3}{\bibinfo {journal} {Phys. Lett. B}}\
  }%
  \textbf{\bibinfo {volume} {150}},\ \bibinfo {pages} {87 } (\bibinfo {year}
  {1985})%
  \bibAnnoteFile{NoStop}{Rosner85}%
\bibitem{Choudhury84}%
  \BibitemOpen
  \bibfield{author}{%
  \bibinfo {author} {\bibfnamefont{R.~K.}\ \bibnamefont{Choudhury}}, \bibinfo
  {author} {\bibfnamefont{P.~L.}\ \bibnamefont{Gonthier}}, \bibinfo {author}
  {\bibfnamefont{K.}~\bibnamefont{Hagel}}, \bibinfo {author}
  {\bibfnamefont{M.~N.}\ \bibnamefont{Namboodiri}}, \bibinfo {author}
  {\bibfnamefont{J.~B.}\ \bibnamefont{Natowitz}}, \bibinfo {author}
  {\bibfnamefont{L.}~\bibnamefont{Adler}}, \bibinfo {author}
  {\bibfnamefont{S.}~\bibnamefont{Simon}}, \bibinfo {author}
  {\bibfnamefont{S.}~\bibnamefont{Kniffen}},\ and\ \bibinfo {author}
  {\bibfnamefont{G.}~\bibnamefont{Berkowitz}},\ }%
  \bibfield{journal}{%
  \Doi{DOI: 10.1016/0370-2693(84)90807-4}{\bibinfo {journal} {Phys. Lett. B}}\
  }%
  \textbf{\bibinfo {volume} {143}},\ \bibinfo {pages} {74 } (\bibinfo {year}
  {1984})%
  \bibAnnoteFile{NoStop}{Choudhury84}%
\bibitem{LaRana87}%
  \BibitemOpen
  \bibfield{author}{%
  \bibinfo {author} {\bibfnamefont{G.}~\bibnamefont{La~Rana}}, \bibinfo
  {author} {\bibfnamefont{D.~J.}\ \bibnamefont{Moses}}, \bibinfo {author}
  {\bibfnamefont{W.~E.}\ \bibnamefont{Parker}}, \bibinfo {author}
  {\bibfnamefont{M.}~\bibnamefont{Kaplan}}, \bibinfo {author}
  {\bibfnamefont{D.}~\bibnamefont{Logan}}, \bibinfo {author}
  {\bibfnamefont{R.}~\bibnamefont{Lacey}}, \bibinfo {author}
  {\bibfnamefont{J.~M.}\ \bibnamefont{Alexander}},\ and\ \bibinfo {author}
  {\bibfnamefont{R.~J.}\ \bibnamefont{Welberry}},\ }%
  \bibfield{journal}{%
  \Doi{10.1103/PhysRevC.35.373}{\bibinfo {journal} {Phys. Rev. C}}\ }%
  \textbf{\bibinfo {volume} {35}},\ \bibinfo {pages} {373} (\bibinfo {year}
  {1987})%
  \bibAnnoteFile{NoStop}{LaRana87}%
\bibitem{Brown99}%
  \BibitemOpen
  \bibfield{author}{%
  \bibinfo {author} {\bibfnamefont{C.~M.}\ \bibnamefont{Brown}}, \bibinfo
  {author} {\bibfnamefont{Z.}~\bibnamefont{Milosevich}}, \bibinfo {author}
  {\bibfnamefont{M.}~\bibnamefont{Kaplan}}, \bibinfo {author}
  {\bibfnamefont{E.}~\bibnamefont{Vardaci}}, \bibinfo {author}
  {\bibfnamefont{P.}~\bibnamefont{DeYoung}}, \bibinfo {author}
  {\bibfnamefont{J.~P.}\ \bibnamefont{Whitfield}}, \bibinfo {author}
  {\bibfnamefont{D.}~\bibnamefont{Peterson}}, \bibinfo {author}
  {\bibfnamefont{C.}~\bibnamefont{Dykstra}}, \bibinfo {author}
  {\bibfnamefont{P.~J.}\ \bibnamefont{Karol}},\ and\ \bibinfo {author}
  {\bibfnamefont{M.~A.}\ \bibnamefont{McMahan}},\ }%
  \bibfield{journal}{%
  \Doi{10.1103/PhysRevC.60.064612}{\bibinfo {journal} {Phys. Rev. C}}\ }%
  \textbf{\bibinfo {volume} {60}},\ \bibinfo {pages} {064612} (\bibinfo {year}
  {1999})%
  \bibAnnoteFile{NoStop}{Brown99}%
\bibitem{Kildir92}%
  \BibitemOpen
  \bibfield{author}{%
  \bibinfo {author} {\bibfnamefont{M.}~\bibnamefont{Kildir}}, \bibinfo {author}
  {\bibfnamefont{G.}~\bibnamefont{La~Rana}}, \bibinfo {author}
  {\bibfnamefont{R.}~\bibnamefont{Moro}}, \bibinfo {author}
  {\bibfnamefont{A.}~\bibnamefont{Brondi}}, \bibinfo {author}
  {\bibfnamefont{A.}~\bibnamefont{D'Onofrio}}, \bibinfo {author}
  {\bibfnamefont{E.}~\bibnamefont{Perillo}}, \bibinfo {author}
  {\bibfnamefont{V.}~\bibnamefont{Roca}}, \bibinfo {author}
  {\bibfnamefont{M.}~\bibnamefont{Romano}}, \bibinfo {author}
  {\bibfnamefont{F.}~\bibnamefont{Terrasi}}, \bibinfo {author}
  {\bibfnamefont{G.}~\bibnamefont{Nebbia}}, \bibinfo {author}
  {\bibfnamefont{G.}~\bibnamefont{Viesti}},\ and\ \bibinfo {author}
  {\bibfnamefont{G.}~\bibnamefont{Prete}},\ }%
  \bibfield{journal}{%
  \Doi{10.1103/PhysRevC.46.2264}{\bibinfo {journal} {Phys. Rev. C}}\ }%
  \textbf{\bibinfo {volume} {46}},\ \bibinfo {pages} {2264} (\bibinfo {year}
  {1992})%
  \bibAnnoteFile{NoStop}{Kildir92}%
\bibitem{Nebbia94}%
  \BibitemOpen
  \bibfield{author}{%
  \bibinfo {author} {\bibfnamefont{G.}~\bibnamefont{Nebbia}}, \bibinfo {author}
  {\bibfnamefont{D.}~\bibnamefont{Fabris}}, \bibinfo {author}
  {\bibfnamefont{A.}~\bibnamefont{Perin}}, \bibinfo {author}
  {\bibfnamefont{G.}~\bibnamefont{Viesti}}, \bibinfo {author}
  {\bibfnamefont{F.}~\bibnamefont{Gamegna}}, \bibinfo {author}
  {\bibfnamefont{G.}~\bibnamefont{Prete}}, \bibinfo {author}
  {\bibfnamefont{L.}~\bibnamefont{Fiore}}, \bibinfo {author}
  {\bibfnamefont{V.}~\bibnamefont{Paticchio}}, \bibinfo {author}
  {\bibfnamefont{F.}~\bibnamefont{Lucarelli}}, \bibinfo {author}
  {\bibfnamefont{B.}~\bibnamefont{Chambon}}, \bibinfo {author}
  {\bibfnamefont{B.}~\bibnamefont{Cheynis}}, \bibinfo {author}
  {\bibfnamefont{D.}~\bibnamefont{Drain}}, \bibinfo {author}
  {\bibfnamefont{A.}~\bibnamefont{Giorni}}, \bibinfo {author}
  {\bibfnamefont{A.}~\bibnamefont{Lleres}},\ and\ \bibinfo {author}
  {\bibfnamefont{J.~B.}\ \bibnamefont{Viano}},\ }%
  \bibfield{journal}{%
  \bibinfo {journal} {Nucl. Phys.}\ }%
  \textbf{\bibinfo {volume} {A578}},\ \bibinfo {pages} {285} (\bibinfo {year}
  {1994})%
  \bibAnnoteFile{NoStop}{Nebbia94}%
\bibitem{Galin74}%
  \BibitemOpen
  \bibfield{author}{%
  \bibinfo {author} {\bibfnamefont{J.}~\bibnamefont{Galin}}, \bibinfo {author}
  {\bibfnamefont{B.}~\bibnamefont{Gatty}}, \bibinfo {author}
  {\bibfnamefont{D.}~\bibnamefont{Guerreau}}, \bibinfo {author}
  {\bibfnamefont{C.}~\bibnamefont{Rousset}}, \bibinfo {author}
  {\bibfnamefont{U.~C.}\ \bibnamefont{Schlotthauer-Voos}},\ and\ \bibinfo
  {author} {\bibfnamefont{X.}~\bibnamefont{Tarrago}},\ }%
  \bibfield{journal}{%
  \Doi{10.1103/PhysRevC.9.1126}{\bibinfo {journal} {Phys. Rev. C}}\ }%
  \textbf{\bibinfo {volume} {9}},\ \bibinfo {pages} {1126} (\bibinfo {year}
  {1974})%
  \bibAnnoteFile{NoStop}{Galin74}%
\bibitem{Galin74a}%
  \BibitemOpen
  \bibfield{author}{%
  \bibinfo {author} {\bibfnamefont{J.}~\bibnamefont{Galin}}, \bibinfo {author}
  {\bibfnamefont{B.}~\bibnamefont{Gatty}}, \bibinfo {author}
  {\bibfnamefont{D.}~\bibnamefont{Guerreau}}, \bibinfo {author}
  {\bibfnamefont{C.}~\bibnamefont{Rousset}}, \bibinfo {author}
  {\bibfnamefont{U.~C.}\ \bibnamefont{Schlottauer-Voos}},\ and\ \bibinfo
  {author} {\bibfnamefont{X.}~\bibnamefont{Tarrago}},\ }%
  \bibfield{journal}{%
  \Doi{10.1103/PhysRevC.9.1113}{\bibinfo {journal} {Phys. Rev. C}}\ }%
  \textbf{\bibinfo {volume} {9}},\ \bibinfo {pages} {1113} (\bibinfo {year}
  {1974})%
  \bibAnnoteFile{NoStop}{Galin74a}%
\bibitem{Janssens86}%
  \BibitemOpen
  \bibfield{author}{%
  \bibinfo {author} {\bibfnamefont{R.~V.~F.}\ \bibnamefont{Janssens}}, \bibinfo
  {author} {\bibfnamefont{R.}~\bibnamefont{Holzmann}}, \bibinfo {author}
  {\bibfnamefont{W.}~\bibnamefont{Henning}}, \bibinfo {author}
  {\bibfnamefont{T.~L.}\ \bibnamefont{Khoo}}, \bibinfo {author}
  {\bibfnamefont{K.~T.}\ \bibnamefont{Lesko}}, \bibinfo {author}
  {\bibfnamefont{G.~S.~F.}\ \bibnamefont{Stephans}}, \bibinfo {author}
  {\bibfnamefont{D.~C.}\ \bibnamefont{Radford}}, \bibinfo {author}
  {\bibfnamefont{A.~M. V.~D.}\ \bibnamefont{Berg}}, \bibinfo {author}
  {\bibfnamefont{W.}~\bibnamefont{K{\"u}hn}},\ and\ \bibinfo {author}
  {\bibfnamefont{R.~M.}\ \bibnamefont{Ronningen}},\ }%
  \bibfield{journal}{%
  \Doi{DOI: 10.1016/0370-2693(86)91245-1}{\bibinfo {journal} {Phys. Lett. B}}\
  }%
  \textbf{\bibinfo {volume} {181}},\ \bibinfo {pages} {16 } (\bibinfo {year}
  {1986})%
  \bibAnnoteFile{NoStop}{Janssens86}%
\bibitem{Charity03}%
  \BibitemOpen
  \bibfield{author}{%
  \bibinfo {author} {\bibfnamefont{R.~J.}\ \bibnamefont{Charity}}, \bibinfo
  {author} {\bibfnamefont{L.~G.}\ \bibnamefont{Sobotka}}, \bibinfo {author}
  {\bibfnamefont{J.~F.}\ \bibnamefont{Dempsey}}, \bibinfo {author}
  {\bibfnamefont{M.}~\bibnamefont{Devlin}}, \bibinfo {author}
  {\bibfnamefont{S.}~\bibnamefont{Komarov}}, \bibinfo {author}
  {\bibfnamefont{D.~G.}\ \bibnamefont{Sarantites}}, \bibinfo {author}
  {\bibfnamefont{A.~L.}\ \bibnamefont{Caraley}}, \bibinfo {author}
  {\bibfnamefont{R.~T.}\ \bibnamefont{deSouza}}, \bibinfo {author}
  {\bibfnamefont{W.}~\bibnamefont{Loveland}}, \bibinfo {author}
  {\bibfnamefont{D.}~\bibnamefont{Peterson}}, \bibinfo {author}
  {\bibfnamefont{B.~B.}\ \bibnamefont{Back}}, \bibinfo {author}
  {\bibfnamefont{C.~N.}\ \bibnamefont{Davids}},\ and\ \bibinfo {author}
  {\bibfnamefont{D.}~\bibnamefont{Seweryniak}},\ }%
  \bibfield{journal}{%
  \Doi{10.1103/PhysRevC.67.044611}{\bibinfo {journal} {Phys. Rev. C}}\ }%
  \textbf{\bibinfo {volume} {67}},\ \bibinfo {pages} {044611} (\bibinfo {year}
  {2003})%
  \bibAnnoteFile{NoStop}{Charity03}%
\bibitem{Fineman94}%
  \BibitemOpen
  \bibfield{author}{%
  \bibinfo {author} {\bibfnamefont{B.~J.}\ \bibnamefont{Fineman}}, \bibinfo
  {author} {\bibfnamefont{K.-T.}\ \bibnamefont{Brinkmann}}, \bibinfo {author}
  {\bibfnamefont{A.~L.}\ \bibnamefont{Caraley}}, \bibinfo {author}
  {\bibfnamefont{N.}~\bibnamefont{Gan}}, \bibinfo {author}
  {\bibfnamefont{R.~L.}\ \bibnamefont{McGrath}},\ and\ \bibinfo {author}
  {\bibfnamefont{J.}~\bibnamefont{Velkovska}},\ }%
  \bibfield{journal}{%
  \Doi{10.1103/PhysRevC.50.1991}{\bibinfo {journal} {Phys. Rev. C}}\ }%
  \textbf{\bibinfo {volume} {50}},\ \bibinfo {pages} {1991} (\bibinfo {year}
  {1994})%
  \bibAnnoteFile{NoStop}{Fineman94}%
\bibitem{Caraley00}%
  \BibitemOpen
  \bibfield{author}{%
  \bibinfo {author} {\bibfnamefont{A.~L.}\ \bibnamefont{Caraley}}, \bibinfo
  {author} {\bibfnamefont{B.~P.}\ \bibnamefont{Henry}}, \bibinfo {author}
  {\bibfnamefont{J.~P.}\ \bibnamefont{Lestone}},\ and\ \bibinfo {author}
  {\bibfnamefont{R.}~\bibnamefont{Vandenbosch}},\ }%
  \bibfield{journal}{%
  \Doi{10.1103/PhysRevC.62.054612}{\bibinfo {journal} {Phys. Rev. C}}\ }%
  \textbf{\bibinfo {volume} {62}},\ \bibinfo {pages} {054612} (\bibinfo {year}
  {2000})%
  \bibAnnoteFile{NoStop}{Caraley00}%
\bibitem{Videbaek77}%
  \BibitemOpen
  \bibfield{author}{%
  \bibinfo {author} {\bibfnamefont{F.}~\bibnamefont{Videb{\ae}k}}, \bibinfo
  {author} {\bibfnamefont{R.~B.}\ \bibnamefont{Goldstein}}, \bibinfo {author}
  {\bibfnamefont{L.}~\bibnamefont{Grodzins}}, \bibinfo {author}
  {\bibfnamefont{S.~G.}\ \bibnamefont{Steadman}}, \bibinfo {author}
  {\bibfnamefont{T.~A.}\ \bibnamefont{Belote}},\ and\ \bibinfo {author}
  {\bibfnamefont{J.~D.}\ \bibnamefont{Garrett}},\ }%
  \bibfield{journal}{%
  \Doi{10.1103/PhysRevC.15.954}{\bibinfo {journal} {Phys. Rev. C}}\ }%
  \textbf{\bibinfo {volume} {15}},\ \bibinfo {pages} {954} (\bibinfo {year}
  {1977})%
  \bibAnnoteFile{NoStop}{Videbaek77}%
\bibitem{Brinkmann94}%
  \BibitemOpen
  \bibfield{author}{%
  \bibinfo {author} {\bibfnamefont{K.-T.}\ \bibnamefont{Brinkmann}}, \bibinfo
  {author} {\bibfnamefont{A.~L.}\ \bibnamefont{Caraley}}, \bibinfo {author}
  {\bibfnamefont{B.~J.}\ \bibnamefont{Fineman}}, \bibinfo {author}
  {\bibfnamefont{N.}~\bibnamefont{Gan}}, \bibinfo {author}
  {\bibfnamefont{J.}~\bibnamefont{Velkovska}},\ and\ \bibinfo {author}
  {\bibfnamefont{R.~L.}\ \bibnamefont{McGrath}},\ }%
  \bibfield{journal}{%
  \Doi{10.1103/PhysRevC.50.309}{\bibinfo {journal} {Phys. Rev. C}}\ }%
  \textbf{\bibinfo {volume} {50}},\ \bibinfo {pages} {309} (\bibinfo {year}
  {1994})%
  \bibAnnoteFile{NoStop}{Brinkmann94}%
\bibitem{Back85}%
  \BibitemOpen
  \bibfield{author}{%
  \bibinfo {author} {\bibfnamefont{B.~B.}\ \bibnamefont{Back}}, \bibinfo
  {author} {\bibfnamefont{R.~R.}\ \bibnamefont{Betts}}, \bibinfo {author}
  {\bibfnamefont{J.~E.}\ \bibnamefont{Gindler}}, \bibinfo {author}
  {\bibfnamefont{B.~D.}\ \bibnamefont{Wilkins}}, \bibinfo {author}
  {\bibfnamefont{S.}~\bibnamefont{Saini}}, \bibinfo {author}
  {\bibfnamefont{M.~B.}\ \bibnamefont{Tsang}}, \bibinfo {author}
  {\bibfnamefont{C.~K.}\ \bibnamefont{Gelbke}}, \bibinfo {author}
  {\bibfnamefont{W.~G.}\ \bibnamefont{Lynch}}, \bibinfo {author}
  {\bibfnamefont{M.~A.}\ \bibnamefont{McMahan}},\ and\ \bibinfo {author}
  {\bibfnamefont{P.~A.}\ \bibnamefont{Baisden}},\ }%
  \bibfield{journal}{%
  \Doi{10.1103/PhysRevC.32.195}{\bibinfo {journal} {Phys. Rev. C}}\ }%
  \textbf{\bibinfo {volume} {32}},\ \bibinfo {pages} {195} (\bibinfo {year}
  {1985})%
  \bibAnnoteFile{NoStop}{Back85}%
\bibitem{Frobrich84}%
  \BibitemOpen
  \bibfield{author}{%
  \bibinfo {author} {\bibfnamefont{P.}~\bibnamefont{Fr{\"o}brich}},\ }%
  \bibfield{journal}{%
  \bibinfo {journal} {Phys. Rep.}\ }%
  \textbf{\bibinfo {volume} {116}},\ \bibinfo {pages} {337} (\bibinfo {year}
  {1984})%
  \bibAnnoteFile{NoStop}{Frobrich84}%
\bibitem{Hauser52}%
  \BibitemOpen
  \bibfield{author}{%
  \bibinfo {author} {\bibfnamefont{W.}~\bibnamefont{Hauser}}\ and\ \bibinfo
  {author} {\bibfnamefont{H.}~\bibnamefont{Feshbach}},\ }%
  \bibfield{journal}{%
  \Doi{10.1103/PhysRev.87.366}{\bibinfo {journal} {Phys. Rev.}}\ }%
  \textbf{\bibinfo {volume} {87}},\ \bibinfo {pages} {366} (\bibinfo {year}
  {1952})%
  \bibAnnoteFile{NoStop}{Hauser52}%
\bibitem{Moretto75}%
  \BibitemOpen
  \bibfield{author}{%
  \bibinfo {author} {\bibfnamefont{L.~G.}\ \bibnamefont{Moretto}},\ }%
  \bibfield{journal}{%
  \bibinfo {journal} {Nucl. Phys.}\ }%
  \textbf{\bibinfo {volume} {A247}},\ \bibinfo {pages} {211} (\bibinfo {year}
  {1975})%
  \bibAnnoteFile{NoStop}{Moretto75}%
\bibitem{Bohr75}%
  \BibitemOpen
  \bibfield{author}{%
  \bibinfo {author} {\bibfnamefont{A.}~\bibnamefont{Bohr}}\ and\ \bibinfo
  {author} {\bibfnamefont{B.~R.}\ \bibnamefont{Mottleson}},\ }%
  \emph{\bibinfo {title} {Nuclear Structure}},\ Vol.~\bibinfo {volume} {I}\
  (\bibinfo {publisher} {Benjamin},\ \bibinfo {address} {New York},\ \bibinfo
  {year} {1975})%
  \bibAnnoteFile{NoStop}{Bohr75}%
\bibitem{Cohen74}%
  \BibitemOpen
  \bibfield{author}{%
  \bibinfo {author} {\bibfnamefont{S.}~\bibnamefont{Cohen}}, \bibinfo {author}
  {\bibfnamefont{F.}~\bibnamefont{Plasil}},\ and\ \bibinfo {author}
  {\bibfnamefont{W.~J.}\ \bibnamefont{Swiatecki}},\ }%
  \bibfield{journal}{%
  \bibinfo {journal} {Ann. Phys. (N.Y.)}\ }%
  \textbf{\bibinfo {volume} {82}},\ \bibinfo {pages} {557} (\bibinfo {year}
  {1974})%
  \bibAnnoteFile{NoStop}{Cohen74}%
\bibitem{Sierk86}%
  \BibitemOpen
  \bibfield{author}{%
  \bibinfo {author} {\bibfnamefont{A.~J.}\ \bibnamefont{Sierk}},\ }%
  \bibfield{journal}{%
  \Doi{10.1103/PhysRevC.33.2039}{\bibinfo {journal} {Phys. Rev. C}}\ }%
  \textbf{\bibinfo {volume} {33}},\ \bibinfo {pages} {2039} (\bibinfo {year}
  {1986})%
  \bibAnnoteFile{NoStop}{Sierk86}%
\bibitem{Alexander90}%
  \BibitemOpen
  \bibfield{author}{%
  \bibinfo {author} {\bibfnamefont{J.~M.}\ \bibnamefont{Alexander}}, \bibinfo
  {author} {\bibfnamefont{M.~T.}\ \bibnamefont{Magda}},\ and\ \bibinfo {author}
  {\bibfnamefont{S.}~\bibnamefont{Landowne}},\ }%
  \bibfield{journal}{%
  \Doi{10.1103/PhysRevC.42.1092}{\bibinfo {journal} {Phys. Rev. C}}\ }%
  \textbf{\bibinfo {volume} {42}},\ \bibinfo {pages} {1092} (\bibinfo {year}
  {1990})%
  \bibAnnoteFile{NoStop}{Alexander90}%
\bibitem{Rawitscher66}%
  \BibitemOpen
  \bibfield{author}{%
  \bibinfo {author} {\bibfnamefont{G.~H.}\ \bibnamefont{Rawitscher}},\ }%
  \bibfield{journal}{%
  \bibinfo {journal} {Nucl. Phys.}\ }%
  \textbf{\bibinfo {volume} {85}},\ \bibinfo {pages} {337} (\bibinfo {year}
  {1966})%
  \bibAnnoteFile{NoStop}{Rawitscher66}%
\bibitem{Perey62}%
  \BibitemOpen
  \bibfield{author}{%
  \bibinfo {author} {\bibfnamefont{C.~M.}\ \bibnamefont{Perey}}\ and\ \bibinfo
  {author} {\bibfnamefont{F.~G.}\ \bibnamefont{Perey}},\ }%
  \bibfield{journal}{%
  \Doi{10.1103/PhysRev.132.755}{\bibinfo {journal} {Phys. Rev.}}\ }%
  \textbf{\bibinfo {volume} {132}},\ \bibinfo {pages} {755} (\bibinfo {year}
  {1963})%
  \bibAnnoteFile{NoStop}{Perey62}%
\bibitem{Perey63}%
  \BibitemOpen
  \bibfield{author}{%
  \bibinfo {author} {\bibfnamefont{F.~G.}\ \bibnamefont{Perey}},\ }%
  \bibfield{journal}{%
  \Doi{10.1103/PhysRev.131.745}{\bibinfo {journal} {Phys. Rev.}}\ }%
  \textbf{\bibinfo {volume} {131}},\ \bibinfo {pages} {745} (\bibinfo {year}
  {1963})%
  \bibAnnoteFile{NoStop}{Perey63}%
\bibitem{Wilmore64}%
  \BibitemOpen
  \bibfield{author}{%
  \bibinfo {author} {\bibfnamefont{D.}~\bibnamefont{Wilmore}}\ and\ \bibinfo
  {author} {\bibfnamefont{P.~E.}\ \bibnamefont{Hodgson}},\ }%
  \bibfield{journal}{%
  \bibinfo {journal} {Nucl. Phys.}\ }%
  \textbf{\bibinfo {volume} {55}},\ \bibinfo {pages} {673} (\bibinfo {year}
  {1964})%
  \bibAnnoteFile{NoStop}{Wilmore64}%
\bibitem{McFadden66}%
  \BibitemOpen
  \bibfield{author}{%
  \bibinfo {author} {\bibfnamefont{L.}~\bibnamefont{{McFadden}}}\ and\ \bibinfo
  {author} {\bibfnamefont{G.~R.}\ \bibnamefont{Satchler}},\ }%
  \bibfield{journal}{%
  \bibinfo {journal} {Nucl. Phys.}\ }%
  \textbf{\bibinfo {volume} {84}},\ \bibinfo {pages} {177} (\bibinfo {year}
  {1966})%
  \bibAnnoteFile{NoStop}{McFadden66}%
\bibitem{Becchetti71}%
  \BibitemOpen
  \bibfield{author}{%
  \bibinfo {author} {\bibfnamefont{F.~D.}\ \bibnamefont{{Becchetti, Jr}}}\ and\
  \bibinfo {author} {\bibfnamefont{G.~W.}\ \bibnamefont{Greenlees}},\ }%
  \enquote{\bibinfo {title} {Polarization phenomena in nuclear reactions},}\ \
  (\bibinfo {publisher} {University of Wisconsin Press, Madison},\ \bibinfo
  {year} {1971})%
  \bibAnnoteFile{NoStop}{Becchetti71}%
\bibitem{Cook82}%
  \BibitemOpen
  \bibfield{author}{%
  \bibinfo {author} {\bibfnamefont{J.}~\bibnamefont{Cook}},\ }%
  \bibfield{journal}{%
  \bibinfo {journal} {Nucl. Phys.}\ }%
  \textbf{\bibinfo {volume} {A388}},\ \bibinfo {pages} {153} (\bibinfo {year}
  {1982})%
  \bibAnnoteFile{NoStop}{Cook82}%
\bibitem{Balzer77}%
  \BibitemOpen
  \bibfield{author}{%
  \bibinfo {author} {\bibfnamefont{R.}~\bibnamefont{Balzer}}, \bibinfo {author}
  {\bibfnamefont{M.}~\bibnamefont{Hugi}}, \bibinfo {author}
  {\bibfnamefont{B.}~\bibnamefont{Kamys}}, \bibinfo {author}
  {\bibfnamefont{J.}~\bibnamefont{Lang}}, \bibinfo {author}
  {\bibfnamefont{R.}~\bibnamefont{M{\"u}ller}}, \bibinfo {author}
  {\bibfnamefont{E.}~\bibnamefont{Ungricht}}, \bibinfo {author}
  {\bibfnamefont{J.}~\bibnamefont{Untern{\"a}hrer}}, \bibinfo {author}
  {\bibfnamefont{L.}~\bibnamefont{Jarczyk}},\ and\ \bibinfo {author}
  {\bibfnamefont{A.}~\bibnamefont{Strza{\l}kowski}},\ }%
  \bibfield{journal}{%
  \bibinfo {journal} {Nucl. Phys.}\ }%
  \textbf{\bibinfo {volume} {A293}},\ \bibinfo {pages} {518} (\bibinfo {year}
  {1977})%
  \bibAnnoteFile{NoStop}{Balzer77}%
\bibitem{Kildir95}%
  \BibitemOpen
  \bibfield{author}{%
  \bibinfo {author} {\bibfnamefont{M.}~\bibnamefont{Kildir}}, \bibinfo {author}
  {\bibfnamefont{G.}~\bibnamefont{La~Rana}}, \bibinfo {author}
  {\bibfnamefont{R.}~\bibnamefont{Moro}}, \bibinfo {author}
  {\bibfnamefont{A.}~\bibnamefont{Brondi}}, \bibinfo {author}
  {\bibfnamefont{E.}~\bibnamefont{Vardaci}}, \bibinfo {author}
  {\bibfnamefont{A.}~\bibnamefont{D'Onofrio}}, \bibinfo {author}
  {\bibfnamefont{D.}~\bibnamefont{Fessas}}, \bibinfo {author}
  {\bibfnamefont{E.}~\bibnamefont{Perillo}}, \bibinfo {author}
  {\bibfnamefont{V.}~\bibnamefont{Roca}}, \bibinfo {author}
  {\bibfnamefont{M.}~\bibnamefont{Romano}}, \bibinfo {author}
  {\bibfnamefont{F.}~\bibnamefont{Terrasi}}, \bibinfo {author}
  {\bibfnamefont{G.}~\bibnamefont{Nebbia}}, \bibinfo {author}
  {\bibfnamefont{G.}~\bibnamefont{Viesti}},\ and\ \bibinfo {author}
  {\bibfnamefont{G.}~\bibnamefont{Prete}},\ }%
  \bibfield{journal}{%
  \Doi{10.1103/PhysRevC.51.1873}{\bibinfo {journal} {Phys. Rev. C}}\ }%
  \textbf{\bibinfo {volume} {51}},\ \bibinfo {pages} {1873} (\bibinfo {year}
  {1995})%
  \bibAnnoteFile{NoStop}{Kildir95}%
\bibitem{Nicolis90}%
  \BibitemOpen
  \bibfield{author}{%
  \bibinfo {author} {\bibfnamefont{N.~G.}\ \bibnamefont{Nicolis}}, \bibinfo
  {author} {\bibfnamefont{D.~G.}\ \bibnamefont{Sarantites}}, \bibinfo {author}
  {\bibfnamefont{L.~A.}\ \bibnamefont{Adler}}, \bibinfo {author}
  {\bibfnamefont{F.~A.}\ \bibnamefont{Dilmanian}}, \bibinfo {author}
  {\bibfnamefont{K.}~\bibnamefont{Honkanen}}, \bibinfo {author}
  {\bibfnamefont{Z.}~\bibnamefont{Majka}}, \bibinfo {author}
  {\bibfnamefont{L.~G.}\ \bibnamefont{Sobotka}}, \bibinfo {author}
  {\bibfnamefont{Z.}~\bibnamefont{Li}}, \bibinfo {author}
  {\bibfnamefont{T.~M.}\ \bibnamefont{Semkow}}, \bibinfo {author}
  {\bibfnamefont{J.~R.}\ \bibnamefont{Beene}}, \bibinfo {author}
  {\bibfnamefont{M.~L.}\ \bibnamefont{Halbert}}, \bibinfo {author}
  {\bibfnamefont{D.~C.}\ \bibnamefont{Hensley}}, \bibinfo {author}
  {\bibfnamefont{J.~B.}\ \bibnamefont{Natowitz}}, \bibinfo {author}
  {\bibfnamefont{R.~P.}\ \bibnamefont{Schmitt}}, \bibinfo {author}
  {\bibfnamefont{D.}~\bibnamefont{Fabris}}, \bibinfo {author}
  {\bibfnamefont{G.}~\bibnamefont{Nebbia}},\ and\ \bibinfo {author}
  {\bibfnamefont{G.}~\bibnamefont{Mouchaty}},\ }%
  \bibfield{journal}{%
  \Doi{10.1103/PhysRevC.41.2118}{\bibinfo {journal} {Phys. Rev. C}}\ }%
  \textbf{\bibinfo {volume} {41}},\ \bibinfo {pages} {2118} (\bibinfo {year}
  {1990})%
  \bibAnnoteFile{NoStop}{Nicolis90}%
\bibitem{Gonin90}%
  \BibitemOpen
  \bibfield{author}{%
  \bibinfo {author} {\bibfnamefont{M.}~\bibnamefont{Gonin}}, \bibinfo {author}
  {\bibfnamefont{L.}~\bibnamefont{Cooke}}, \bibinfo {author}
  {\bibfnamefont{K.}~\bibnamefont{Hagel}}, \bibinfo {author}
  {\bibfnamefont{Y.}~\bibnamefont{Lou}}, \bibinfo {author}
  {\bibfnamefont{J.~B.}\ \bibnamefont{Natowitz}}, \bibinfo {author}
  {\bibfnamefont{R.~P.}\ \bibnamefont{Schmitt}}, \bibinfo {author}
  {\bibfnamefont{S.}~\bibnamefont{Shlomo}}, \bibinfo {author}
  {\bibfnamefont{B.}~\bibnamefont{Srivastava}}, \bibinfo {author}
  {\bibfnamefont{W.}~\bibnamefont{Turmel}}, \bibinfo {author}
  {\bibfnamefont{H.}~\bibnamefont{Utsunomiya}}, \bibinfo {author}
  {\bibfnamefont{R.}~\bibnamefont{Wada}}, \bibinfo {author}
  {\bibfnamefont{G.}~\bibnamefont{Nardelli}}, \bibinfo {author}
  {\bibfnamefont{G.}~\bibnamefont{Nebbia}}, \bibinfo {author}
  {\bibfnamefont{G.}~\bibnamefont{Viesti}}, \bibinfo {author}
  {\bibfnamefont{R.}~\bibnamefont{Zanon}}, \bibinfo {author}
  {\bibfnamefont{B.}~\bibnamefont{Fornal}}, \bibinfo {author}
  {\bibfnamefont{G.}~\bibnamefont{Prete}}, \bibinfo {author}
  {\bibfnamefont{K.}~\bibnamefont{Niita}}, \bibinfo {author}
  {\bibfnamefont{S.}~\bibnamefont{Hannuschke}}, \bibinfo {author}
  {\bibfnamefont{P.}~\bibnamefont{Gonthier}},\ and\ \bibinfo {author}
  {\bibfnamefont{B.}~\bibnamefont{Wilkins}},\ }%
  \bibfield{journal}{%
  \Doi{10.1103/PhysRevC.42.2125}{\bibinfo {journal} {Phys. Rev. C}}\ }%
  \textbf{\bibinfo {volume} {42}},\ \bibinfo {pages} {2125} (\bibinfo {year}
  {1990})%
  \bibAnnoteFile{NoStop}{Gonin90}%
\bibitem{Boger94}%
  \BibitemOpen
  \bibfield{author}{%
  \bibinfo {author} {\bibfnamefont{J.}~\bibnamefont{Boger}}, \bibinfo {author}
  {\bibfnamefont{J.~M.}\ \bibnamefont{Alexander}}, \bibinfo {author}
  {\bibfnamefont{R.~A.}\ \bibnamefont{Lacey}},\ and\ \bibinfo {author}
  {\bibfnamefont{A.}~\bibnamefont{Narayanan}},\ }%
  \bibfield{journal}{%
  \Doi{10.1103/PhysRevC.49.1587}{\bibinfo {journal} {Phys. Rev. C}}\ }%
  \textbf{\bibinfo {volume} {49}},\ \bibinfo {pages} {1587} (\bibinfo {year}
  {1994})%
  \bibAnnoteFile{NoStop}{Boger94}%
\bibitem{Lacey87}%
  \BibitemOpen
  \bibfield{author}{%
  \bibinfo {author} {\bibfnamefont{R.}~\bibnamefont{Lacey}}, \bibinfo {author}
  {\bibfnamefont{N.~N.}\ \bibnamefont{Ajitanand}}, \bibinfo {author}
  {\bibfnamefont{J.~M.}\ \bibnamefont{Alexander}}, \bibinfo {author}
  {\bibfnamefont{D.~M. D.~C.}\ \bibnamefont{Rizzo}}, \bibinfo {author}
  {\bibfnamefont{P.}~\bibnamefont{Deyoung}}, \bibinfo {author}
  {\bibfnamefont{M.}~\bibnamefont{Kaplan}}, \bibinfo {author}
  {\bibfnamefont{L.}~\bibnamefont{Kowalski}}, \bibinfo {author}
  {\bibfnamefont{G.~L.}\ \bibnamefont{Rana}}, \bibinfo {author}
  {\bibfnamefont{D.}~\bibnamefont{Logan}}, \bibinfo {author}
  {\bibfnamefont{D.~J.}\ \bibnamefont{Moses}}, \bibinfo {author}
  {\bibfnamefont{W.~E.}\ \bibnamefont{Parker}}, \bibinfo {author}
  {\bibfnamefont{G.~F.}\ \bibnamefont{Peaslee}},\ and\ \bibinfo {author}
  {\bibfnamefont{L.~C.}\ \bibnamefont{Vaz}},\ }%
  \bibfield{journal}{%
  \Doi{DOI: 10.1016/0370-2693(87)90250-4}{\bibinfo {journal} {Phys. Lett. B}}\
  }%
  \textbf{\bibinfo {volume} {191}},\ \bibinfo {pages} {253 } (\bibinfo {year}
  {1987})%
  \bibAnnoteFile{NoStop}{Lacey87}%
\bibitem{Charity01}%
  \BibitemOpen
  \bibfield{author}{%
  \bibinfo {author} {\bibfnamefont{R.~J.}\ \bibnamefont{Charity}}, \bibinfo
  {author} {\bibfnamefont{L.~G.}\ \bibnamefont{Sobotka}}, \bibinfo {author}
  {\bibfnamefont{J.}~\bibnamefont{Cibor}}, \bibinfo {author}
  {\bibfnamefont{K.}~\bibnamefont{Hagel}}, \bibinfo {author}
  {\bibfnamefont{M.}~\bibnamefont{Murray}}, \bibinfo {author}
  {\bibfnamefont{J.~B.}\ \bibnamefont{Natowitz}}, \bibinfo {author}
  {\bibfnamefont{R.}~\bibnamefont{Wada}}, \bibinfo {author}
  {\bibfnamefont{Y.}~\bibnamefont{El~Masri}}, \bibinfo {author}
  {\bibfnamefont{D.}~\bibnamefont{Fabris}}, \bibinfo {author}
  {\bibfnamefont{G.}~\bibnamefont{Nebbia}}, \bibinfo {author}
  {\bibfnamefont{G.}~\bibnamefont{Viesti}}, \bibinfo {author}
  {\bibfnamefont{M.}~\bibnamefont{Cinausero}}, \bibinfo {author}
  {\bibfnamefont{E.}~\bibnamefont{Fioretto}}, \bibinfo {author}
  {\bibfnamefont{G.}~\bibnamefont{Prete}}, \bibinfo {author}
  {\bibfnamefont{A.}~\bibnamefont{Wagner}},\ and\ \bibinfo {author}
  {\bibfnamefont{H.}~\bibnamefont{Xu}},\ }%
  \bibfield{journal}{%
  \Doi{10.1103/PhysRevC.63.024611}{\bibinfo {journal} {Phys. Rev. C}}\ }%
  \textbf{\bibinfo {volume} {63}},\ \bibinfo {pages} {024611} (\bibinfo {year}
  {2001})%
  \bibAnnoteFile{NoStop}{Charity01}%
\bibitem{Huizenga89}%
  \BibitemOpen
  \bibfield{author}{%
  \bibinfo {author} {\bibfnamefont{J.~R.}\ \bibnamefont{Huizenga}}, \bibinfo
  {author} {\bibfnamefont{A.~N.}\ \bibnamefont{Behkami}}, \bibinfo {author}
  {\bibfnamefont{I.~M.}\ \bibnamefont{Govil}}, \bibinfo {author}
  {\bibfnamefont{W.~U.}\ \bibnamefont{Schr{\"o}der}},\ and\ \bibinfo {author}
  {\bibfnamefont{J.}~\bibnamefont{T{\~o}ke}},\ }%
  \bibfield{journal}{%
  \Doi{10.1103/PhysRevC.40.668}{\bibinfo {journal} {Phys. Rev. C}}\ }%
  \textbf{\bibinfo {volume} {40}},\ \bibinfo {pages} {668} (\bibinfo {year}
  {1989})%
  \bibAnnoteFile{NoStop}{Huizenga89}%
\bibitem{Charity00}%
  \BibitemOpen
  \bibfield{author}{%
  \bibinfo {author} {\bibfnamefont{R.~J.}\ \bibnamefont{Charity}},\ }%
  \bibfield{journal}{%
  \Doi{10.1103/PhysRevC.61.054614}{\bibinfo {journal} {Phys. Rev. C}}\ }%
  \textbf{\bibinfo {volume} {61}},\ \bibinfo {pages} {054614} (\bibinfo {year}
  {2000})%
  \bibAnnoteFile{NoStop}{Charity00}%
\bibitem{Charity01a}%
  \BibitemOpen
  \bibfield{author}{%
  \bibinfo {author} {\bibfnamefont{R.~J.}\ \bibnamefont{Charity}},\ }%
  \bibfield{journal}{%
  \Doi{10.1103/PhysRevC.64.064610}{\bibinfo {journal} {Phys. Rev. C}}\ }%
  \textbf{\bibinfo {volume} {64}},\ \bibinfo {pages} {064610} (\bibinfo {year}
  {2001})%
  \bibAnnoteFile{NoStop}{Charity01a}%
\bibitem{Santo63}%
  \BibitemOpen
  \bibfield{author}{%
  \bibinfo {author} {\bibfnamefont{M.}~\bibnamefont{Santo}}\ and\ \bibinfo
  {author} {\bibfnamefont{S.}~\bibnamefont{Yamasaki}},\ }%
  \bibfield{journal}{%
  \bibinfo {journal} {Prog. Theor. Phys.}\ }%
  \textbf{\bibinfo {volume} {29}},\ \bibinfo {pages} {397} (\bibinfo {year}
  {1963})%
  \bibAnnoteFile{NoStop}{Santo63}%
\bibitem{Moretto72}%
  \BibitemOpen
  \bibfield{author}{%
  \bibinfo {author} {\bibfnamefont{L.~G.}\ \bibnamefont{Moretto}},\ }%
  \bibfield{journal}{%
  \Doi{DOI: 10.1016/0375-9474(72)90556-8}{\bibinfo {journal} {Nucl. Phy. A}}\
  }%
  \textbf{\bibinfo {volume} {185}},\ \bibinfo {pages} {145 } (\bibinfo {year}
  {1972})%
  \bibAnnoteFile{NoStop}{Moretto72}%
\bibitem{John01}%
  \BibitemOpen
  \bibfield{author}{%
  \bibinfo {author} {\bibfnamefont{B.}~\bibnamefont{John}}, \bibinfo {author}
  {\bibfnamefont{R.~K.}\ \bibnamefont{Choudhury}}, \bibinfo {author}
  {\bibfnamefont{B.~K.}\ \bibnamefont{Nayak}}, \bibinfo {author}
  {\bibfnamefont{A.}~\bibnamefont{Saxena}},\ and\ \bibinfo {author}
  {\bibfnamefont{D.~C.}\ \bibnamefont{Biswas}},\ }%
  \bibfield{journal}{%
  \Doi{10.1103/PhysRevC.63.054301}{\bibinfo {journal} {Phys. Rev. C}}\ }%
  \textbf{\bibinfo {volume} {63}},\ \bibinfo {pages} {054301} (\bibinfo {year}
  {2001})%
  \bibAnnoteFile{NoStop}{John01}%
\bibitem{Ignatyuk75}%
  \BibitemOpen
  \bibfield{author}{%
  \bibinfo {author} {\bibfnamefont{A.~V.}\ \bibnamefont{Ignatyuk}}, \bibinfo
  {author} {\bibfnamefont{G.~N.}\ \bibnamefont{Smirenkin}},\ and\ \bibinfo
  {author} {\bibfnamefont{A.~S.}\ \bibnamefont{Tishin}},\ }%
  \bibfield{journal}{%
  \bibinfo {journal} {Sov. J. Nucl. Phys.}\ }%
  \textbf{\bibinfo {volume} {21}},\ \bibinfo {pages} {255} (\bibinfo {year}
  {1975})%
  \bibAnnoteFile{NoStop}{Ignatyuk75}%
\bibitem{Moller95}%
  \BibitemOpen
  \bibfield{author}{%
  \bibinfo {author} {\bibfnamefont{P.}~\bibnamefont{M{\"o}ller}}, \bibinfo
  {author} {\bibfnamefont{J.~R.}\ \bibnamefont{Nix}}, \bibinfo {author}
  {\bibfnamefont{W.~D.}\ \bibnamefont{Myers}},\ and\ \bibinfo {author}
  {\bibfnamefont{W.~J.}\ \bibnamefont{Swiatecki}},\ }%
  \bibfield{journal}{%
  \bibinfo {journal} {At. Data Nucl. Data Tables}\ }%
  \textbf{\bibinfo {volume} {59}},\ \bibinfo {pages} {185} (\bibinfo {year}
  {1995})%
  \bibAnnoteFile{NoStop}{Moller95}%
\bibitem{Ignatyuk76}%
  \BibitemOpen
  \bibfield{author}{%
  \bibinfo {author} {\bibfnamefont{A.~V.}\ \bibnamefont{Ignatyuk}}, \bibinfo
  {author} {\bibfnamefont{M.~G.}\ \bibnamefont{Itkis}}, \bibinfo {author}
  {\bibfnamefont{V.~N.}\ \bibnamefont{Okolovich}}, \bibinfo {author}
  {\bibfnamefont{G.~N.}\ \bibnamefont{Smirenkin}},\ and\ \bibinfo {author}
  {\bibfnamefont{A.~S.}\ \bibnamefont{Tishin}},\ }%
  \bibfield{journal}{%
  \bibinfo {journal} {Sov. J. Nucl. Phys.}\ }%
  \textbf{\bibinfo {volume} {21}},\ \bibinfo {pages} {612} (\bibinfo {year}
  {1976})%
  \bibAnnoteFile{NoStop}{Ignatyuk76}%
\bibitem{charity05}%
  \BibitemOpen
  \bibfield{author}{%
  \bibinfo {author} {\bibfnamefont{R.~J.}\ \bibnamefont{Charity}}\ and\
  \bibinfo {author} {\bibfnamefont{L.~G.}\ \bibnamefont{Sobotka}},\ }%
  \bibfield{journal}{%
  \Doi{10.1103/PhysRevC.71.024310}{\bibinfo {journal} {Phys. Rev. C}}\ }%
  \textbf{\bibinfo {volume} {71}},\ \bibinfo {eid} {024310} (\bibinfo {year}
  {2005})%
  \bibAnnoteFile{NoStop}{charity05}%
\bibitem{Bjornholm74}%
  \BibitemOpen
  \bibfield{author}{%
  \bibinfo {author} {\bibfnamefont{S.}~\bibnamefont{Bj{\o}rnholm}}, \bibinfo
  {author} {\bibfnamefont{A.}~\bibnamefont{Bohr}},\ and\ \bibinfo {author}
  {\bibfnamefont{B.~R.}\ \bibnamefont{Mottelson}},\ }%
  in\ \emph{\bibinfo {booktitle} {Proceedings of the International Conference
  on the Physics and Chemistry of Fission, Rochester, New York, 1973}},\
  Vol.~\bibinfo {volume} {1}\ (\bibinfo {publisher} {IAEA},\ \bibinfo {address}
  {Vienna},\ \bibinfo {year} {1974})\ p.\ \bibinfo {pages} {367}%
  \bibAnnoteFile{NoStop}{Bjornholm74}%
\bibitem{Hansen83}%
  \BibitemOpen
  \bibfield{author}{%
  \bibinfo {author} {\bibfnamefont{G.}~\bibnamefont{Hansen}}\ and\ \bibinfo
  {author} {\bibfnamefont{A.~S.}\ \bibnamefont{Jensen}},\ }%
  \bibfield{journal}{%
  \bibinfo {journal} {Nucl. Phys.}\ }%
  \textbf{\bibinfo {volume} {A406}},\ \bibinfo {pages} {236} (\bibinfo {year}
  {1983})%
  \bibAnnoteFile{NoStop}{Hansen83}%
\bibitem{Mahaux91}%
  \BibitemOpen
  \bibfield{author}{%
  \bibinfo {author} {\bibfnamefont{C.}~\bibnamefont{Mahaux}}\ and\ \bibinfo
  {author} {\bibfnamefont{R.}~\bibnamefont{Sartor}},\ }%
  \bibfield{journal}{%
  \bibinfo {journal} {Adv. Nucl. Phys.}\ }%
  \textbf{\bibinfo {volume} {20}},\ \bibinfo {pages} {1} (\bibinfo {year}
  {1991})%
  \bibAnnoteFile{NoStop}{Mahaux91}%
\bibitem{Shlomo91}%
  \BibitemOpen
  \bibfield{author}{%
  \bibinfo {author} {\bibfnamefont{S.}~\bibnamefont{Shlomo}}\ and\ \bibinfo
  {author} {\bibfnamefont{J.~B.}\ \bibnamefont{Natowitz}},\ }%
  \bibfield{journal}{%
  \Doi{10.1103/PhysRevC.44.2878}{\bibinfo {journal} {Phys. Rev. C}}\ }%
  \textbf{\bibinfo {volume} {44}},\ \bibinfo {pages} {2878} (\bibinfo {year}
  {1991})%
  \bibAnnoteFile{NoStop}{Shlomo91}%
\bibitem{Sobotka06}%
  \BibitemOpen
  \bibfield{author}{%
  \bibinfo {author} {\bibfnamefont{L.~G.}\ \bibnamefont{Sobotka}}\ and\
  \bibinfo {author} {\bibfnamefont{R.~J.}\ \bibnamefont{Charity}},\ }%
  \bibfield{journal}{%
  \bibinfo {journal} {Phys. Rev. C}\ }%
  \textbf{\bibinfo {volume} {73}},\ \bibinfo {pages} {014609} (\bibinfo {year}
  {2006})%
  \bibAnnoteFile{NoStop}{Sobotka06}%
\bibitem{Alhassid03}%
  \BibitemOpen
  \bibfield{author}{%
  \bibinfo {author} {\bibfnamefont{Y.}~\bibnamefont{Alhassid}}, \bibinfo
  {author} {\bibfnamefont{G.~F.}\ \bibnamefont{Bertsch}},\ and\ \bibinfo
  {author} {\bibfnamefont{L.}~\bibnamefont{Fang}},\ }%
  \bibfield{journal}{%
  \Doi{10.1103/PhysRevC.68.044322}{\bibinfo {journal} {Phys. Rev. C}}\ }%
  \textbf{\bibinfo {volume} {68}},\ \bibinfo {pages} {044322} (\bibinfo {year}
  {2003})%
  \bibAnnoteFile{NoStop}{Alhassid03}%
\bibitem{Bohr39}%
  \BibitemOpen
  \bibfield{author}{%
  \bibinfo {author} {\bibfnamefont{N.}~\bibnamefont{Bohr}}\ and\ \bibinfo
  {author} {\bibfnamefont{J.~A.}\ \bibnamefont{Wheeler}},\ }%
  \bibfield{journal}{%
  \Doi{10.1103/PhysRev.56.426}{\bibinfo {journal} {Phys. Rev.}}\ }%
  \textbf{\bibinfo {volume} {56}},\ \bibinfo {pages} {426} (\bibinfo {year}
  {1939})%
  \bibAnnoteFile{NoStop}{Bohr39}%
\bibitem{Kramers40}%
  \BibitemOpen
  \bibfield{author}{%
  \bibinfo {author} {\bibfnamefont{H.~A.}\ \bibnamefont{Kramers}},\ }%
  \bibfield{journal}{%
  \bibinfo {journal} {Physica}\ }%
  \textbf{\bibinfo {volume} {7}},\ \bibinfo {pages} {284} (\bibinfo {year}
  {1940})%
  \bibAnnoteFile{NoStop}{Kramers40}%
\bibitem{Grange83}%
  \BibitemOpen
  \bibfield{author}{%
  \bibinfo {author} {\bibfnamefont{P.}~\bibnamefont{Grang{\'e}}}, \bibinfo
  {author} {\bibfnamefont{L.}~\bibnamefont{Jun-Qing}},\ and\ \bibinfo {author}
  {\bibfnamefont{H.~A.}\ \bibnamefont{Weidenm{\"u}ller}},\ }%
  \bibfield{journal}{%
  \Doi{10.1103/PhysRevC.27.2063}{\bibinfo {journal} {Phys. Rev. C}}\ }%
  \textbf{\bibinfo {volume} {27}},\ \bibinfo {pages} {2063} (\bibinfo {year}
  {1983})%
  \bibAnnoteFile{NoStop}{Grange83}%
\bibitem{Hilscher92}%
  \BibitemOpen
  \bibfield{author}{%
  \bibinfo {author} {\bibfnamefont{D.}~\bibnamefont{Hilscher}}\ and\ \bibinfo
  {author} {\bibfnamefont{H.}~\bibnamefont{Rossner}},\ }%
  \bibfield{journal}{%
  \bibinfo {journal} {Ann. Phys. (Paris)}\ }%
  \textbf{\bibinfo {volume} {17}},\ \bibinfo {pages} {471} (\bibinfo {year}
  {1992})%
  \bibAnnoteFile{NoStop}{Hilscher92}%
\bibitem{Frobrich93}%
  \BibitemOpen
  \bibfield{author}{%
  \bibinfo {author} {\bibfnamefont{P.}~\bibnamefont{Fr{\"o}brich}}, \bibinfo
  {author} {\bibfnamefont{I.~I.}\ \bibnamefont{Gontchar}},\ and\ \bibinfo
  {author} {\bibfnamefont{N.~D.}\ \bibnamefont{Mavlitov}},\ }%
  \bibfield{journal}{%
  \Doi{DOI: 10.1016/0375-9474(93)90352-X}{\bibinfo {journal} {Nucl. Phys.}}\ }%
  \textbf{\bibinfo {volume} {A556}},\ \bibinfo {pages} {281 } (\bibinfo {year}
  {1993})%
  \bibAnnoteFile{NoStop}{Frobrich93}%
\bibitem{Gontchar09}%
  \BibitemOpen
  \bibfield{author}{%
  \bibinfo {author} {\bibfnamefont{I.~I.}\ \bibnamefont{Gontchar}}\ and\
  \bibinfo {author} {\bibfnamefont{N.~E.}\ \bibnamefont{Aktaev}},\ }%
  \bibfield{journal}{%
  \Doi{10.1103/PhysRevC.80.044601}{\bibinfo {journal} {Phy. Rev. C}}\ }%
  \textbf{\bibinfo {volume} {80}},\ \bibinfo {eid} {044601} (\bibinfo {year}
  {2009})%
  \bibAnnoteFile{NoStop}{Gontchar09}%
\bibitem{Back99}%
  \BibitemOpen
  \bibfield{author}{%
  \bibinfo {author} {\bibfnamefont{B.~B.}\ \bibnamefont{Back}}, \bibinfo
  {author} {\bibfnamefont{D.~J.}\ \bibnamefont{Blumenthal}}, \bibinfo {author}
  {\bibfnamefont{C.~N.}\ \bibnamefont{Davids}}, \bibinfo {author}
  {\bibfnamefont{D.~J.}\ \bibnamefont{Henderson}}, \bibinfo {author}
  {\bibfnamefont{R.}~\bibnamefont{Hermann}}, \bibinfo {author}
  {\bibfnamefont{D.~J.}\ \bibnamefont{Hofman}}, \bibinfo {author}
  {\bibfnamefont{C.~L.}\ \bibnamefont{Jiang}}, \bibinfo {author}
  {\bibfnamefont{H.~T.}\ \bibnamefont{Penttil{\"a}}},\ and\ \bibinfo {author}
  {\bibfnamefont{A.~H.}\ \bibnamefont{Wuosmaa}},\ }%
  \bibfield{journal}{%
  \Doi{10.1103/PhysRevC.60.044602}{\bibinfo {journal} {Phys. Rev. C}}\ }%
  \textbf{\bibinfo {volume} {60}},\ \bibinfo {pages} {044602} (\bibinfo {year}
  {1999})%
  \bibAnnoteFile{NoStop}{Back99}%
\bibitem{Moretto95}%
  \BibitemOpen
  \bibfield{author}{%
  \bibinfo {author} {\bibfnamefont{L.~G.}\ \bibnamefont{Moretto}}, \bibinfo
  {author} {\bibfnamefont{K.~X.}\ \bibnamefont{Jing}}, \bibinfo {author}
  {\bibfnamefont{R.}~\bibnamefont{Gatti}}, \bibinfo {author}
  {\bibfnamefont{G.~J.}\ \bibnamefont{Wozniak}},\ and\ \bibinfo {author}
  {\bibfnamefont{R.~P.}\ \bibnamefont{Schmitt}},\ }%
  \bibfield{journal}{%
  \Doi{10.1103/PhysRevLett.75.4186}{\bibinfo {journal} {Phys. Rev. Lett.}}\ }%
  \textbf{\bibinfo {volume} {75}},\ \bibinfo {pages} {4186} (\bibinfo {year}
  {1995})%
  \bibAnnoteFile{NoStop}{Moretto95}%
\bibitem{Lestone09}%
  \BibitemOpen
  \bibfield{author}{%
  \bibinfo {author} {\bibfnamefont{J.~P.}\ \bibnamefont{Lestone}}\ and\
  \bibinfo {author} {\bibfnamefont{S.~G.}\ \bibnamefont{McCalla}},\ }%
  \bibfield{journal}{%
  \Doi{10.1103/PhysRevC.79.044611}{\bibinfo {journal} {Phys. Rev. C}}\ }%
  \textbf{\bibinfo {volume} {79}},\ \bibinfo {eid} {044611} (\bibinfo {year}
  {2009})%
  \bibAnnoteFile{NoStop}{Lestone09}%
\bibitem{Tishchenko05}%
  \BibitemOpen
  \bibfield{author}{%
  \bibinfo {author} {\bibfnamefont{V.}~\bibnamefont{Tishchenko}}, \bibinfo
  {author} {\bibfnamefont{C.-M.}\ \bibnamefont{Herbach}}, \bibinfo {author}
  {\bibfnamefont{D.}~\bibnamefont{Hilscher}}, \bibinfo {author}
  {\bibfnamefont{U.}~\bibnamefont{Jahnke}}, \bibinfo {author}
  {\bibfnamefont{J.}~\bibnamefont{Galin}}, \bibinfo {author}
  {\bibfnamefont{F.}~\bibnamefont{Goldenbaum}}, \bibinfo {author}
  {\bibfnamefont{A.}~\bibnamefont{Letourneau}},\ and\ \bibinfo {author}
  {\bibfnamefont{W.-U.}\ \bibnamefont{Schr{\"o}der}},\ }%
  \bibfield{journal}{%
  \Doi{10.1103/PhysRevLett.95.162701}{\bibinfo {journal} {Phys. Rev. Lett.}}\
  }%
  \textbf{\bibinfo {volume} {95}},\ \bibinfo {pages} {162701} (\bibinfo {year}
  {2005})%
  \bibAnnoteFile{NoStop}{Tishchenko05}%
\bibitem{Toke81}%
  \BibitemOpen
  \bibfield{author}{%
  \bibinfo {author} {\bibfnamefont{J.}~\bibnamefont{T{\~o}ke}}\ and\ \bibinfo
  {author} {\bibfnamefont{W.}~\bibnamefont{\'{S}wiatecki}},\ }%
  \bibfield{journal}{%
  \bibinfo {journal} {Nucl. Phys.}\ }%
  \textbf{\bibinfo {volume} {A372}},\ \bibinfo {pages} {141} (\bibinfo {year}
  {1981})%
  \bibAnnoteFile{NoStop}{Toke81}%
\bibitem{Hinde82}%
  \BibitemOpen
  \bibfield{author}{%
  \bibinfo {author} {\bibfnamefont{D.~J.}\ \bibnamefont{Hinde}}, \bibinfo
  {author} {\bibfnamefont{J.~R.}\ \bibnamefont{Leigh}}, \bibinfo {author}
  {\bibfnamefont{J.~O.}\ \bibnamefont{Newton}}, \bibinfo {author}
  {\bibfnamefont{W.}~\bibnamefont{Galster}},\ and\ \bibinfo {author}
  {\bibfnamefont{S.}~\bibnamefont{Sie}},\ }%
  \bibfield{journal}{%
  \bibinfo {journal} {Nucl. Phys.}\ }%
  \textbf{\bibinfo {volume} {A385}},\ \bibinfo {pages} {109} (\bibinfo {year}
  {1982})%
  \bibAnnoteFile{NoStop}{Hinde82}%
\bibitem{Fabris94}%
  \BibitemOpen
  \bibfield{author}{%
  \bibinfo {author} {\bibfnamefont{D.}~\bibnamefont{Fabris}}, \bibinfo {author}
  {\bibfnamefont{E.}~\bibnamefont{Fioretto}}, \bibinfo {author}
  {\bibfnamefont{G.}~\bibnamefont{Viesti}}, \bibinfo {author}
  {\bibfnamefont{M.}~\bibnamefont{Cinausero}}, \bibinfo {author}
  {\bibfnamefont{N.}~\bibnamefont{Gelli}}, \bibinfo {author}
  {\bibfnamefont{K.}~\bibnamefont{Hagel}}, \bibinfo {author}
  {\bibfnamefont{F.}~\bibnamefont{Lucarelli}}, \bibinfo {author}
  {\bibfnamefont{J.~B.}\ \bibnamefont{Natowitz}}, \bibinfo {author}
  {\bibfnamefont{G.}~\bibnamefont{Nebbia}}, \bibinfo {author}
  {\bibfnamefont{G.}~\bibnamefont{Prete}},\ and\ \bibinfo {author}
  {\bibfnamefont{R.}~\bibnamefont{Wada}},\ }%
  \bibfield{journal}{%
  \Doi{10.1103/PhysRevC.50.R1261}{\bibinfo {journal} {Phys. Rev. C}}\ }%
  \textbf{\bibinfo {volume} {50}},\ \bibinfo {pages} {R1261} (\bibinfo {year}
  {1994})%
  \bibAnnoteFile{NoStop}{Fabris94}%
\bibitem{Berriman01}%
  \BibitemOpen
  \bibfield{author}{%
  \bibinfo {author} {\bibfnamefont{A.~C.}\ \bibnamefont{Berriman}}, \bibinfo
  {author} {\bibfnamefont{D.~J.}\ \bibnamefont{Hinde}}, \bibinfo {author}
  {\bibfnamefont{M.}~\bibnamefont{Dasgupt}}, \bibinfo {author}
  {\bibfnamefont{C.~R.}\ \bibnamefont{Morton}}, \bibinfo {author}
  {\bibfnamefont{R.~D.}\ \bibnamefont{Butt}},\ and\ \bibinfo {author}
  {\bibfnamefont{J.~O.}\ \bibnamefont{Newton}},\ }%
  \bibfield{journal}{%
  \bibinfo {journal} {Nature (London)}\ }%
  \textbf{\bibinfo {volume} {413}},\ \bibinfo {pages} {144} (\bibinfo {year}
  {2001})%
  \bibAnnoteFile{NoStop}{Berriman01}%
\bibitem{Keller87}%
  \BibitemOpen
  \bibfield{author}{%
  \bibinfo {author} {\bibfnamefont{J.~G.}\ \bibnamefont{Keller}}, \bibinfo
  {author} {\bibfnamefont{B.~B.}\ \bibnamefont{Back}}, \bibinfo {author}
  {\bibfnamefont{B.~G.}\ \bibnamefont{Glagola}}, \bibinfo {author}
  {\bibfnamefont{D.}~\bibnamefont{Henderson}}, \bibinfo {author}
  {\bibfnamefont{S.~B.}\ \bibnamefont{Kaufman}}, \bibinfo {author}
  {\bibfnamefont{S.~J.}\ \bibnamefont{Sanders}}, \bibinfo {author}
  {\bibfnamefont{R.~H.}\ \bibnamefont{Siemssen}}, \bibinfo {author}
  {\bibfnamefont{F.}~\bibnamefont{Videbaek}}, \bibinfo {author}
  {\bibfnamefont{B.~D.}\ \bibnamefont{Wilkins}},\ and\ \bibinfo {author}
  {\bibfnamefont{A.}~\bibnamefont{Worsham}},\ }%
  \bibfield{journal}{%
  \Doi{10.1103/PhysRevC.36.1364}{\bibinfo {journal} {Phys. Rev. C}}\ }%
  \textbf{\bibinfo {volume} {36}},\ \bibinfo {pages} {1364} (\bibinfo {year}
  {1987})%
  \bibAnnoteFile{NoStop}{Keller87}%
\bibitem{Hinde02}%
  \BibitemOpen
  \bibfield{author}{%
  \bibinfo {author} {\bibfnamefont{R.}~\bibnamefont{Rafiei}}, \bibinfo {author}
  {\bibfnamefont{R.~G.}\ \bibnamefont{Thomas}}, \bibinfo {author}
  {\bibfnamefont{D.~J.}\ \bibnamefont{Hinde}}, \bibinfo {author}
  {\bibfnamefont{M.}~\bibnamefont{Dasgupta}}, \bibinfo {author}
  {\bibfnamefont{C.~R.}\ \bibnamefont{Morton}}, \bibinfo {author}
  {\bibfnamefont{L.~R.}\ \bibnamefont{Gasques}}, \bibinfo {author}
  {\bibfnamefont{M.~L.}\ \bibnamefont{Brown}},\ and\ \bibinfo {author}
  {\bibfnamefont{M.~D.}\ \bibnamefont{Rodriguez}},\ }%
  \bibfield{journal}{%
  \Doi{10.1103/PhysRevC.77.024606}{\bibinfo {journal} {Phys. Rev. C}}\ }%
  \textbf{\bibinfo {volume} {77}},\ \bibinfo {pages} {024606} (\bibinfo {year}
  {2008})%
  \bibAnnoteFile{NoStop}{Hinde02}%
\bibitem{Oganessian04}%
  \BibitemOpen
  \bibfield{author}{%
  \bibinfo {author} {\bibfnamefont{Y.~T.}\ \bibnamefont{Oganessian}}, \bibinfo
  {author} {\bibfnamefont{V.~K.}\ \bibnamefont{Utyonkov}}, \bibinfo {author}
  {\bibfnamefont{Y.~V.}\ \bibnamefont{Lobanov}}, \bibinfo {author}
  {\bibfnamefont{F.~S.}\ \bibnamefont{Abdullin}}, \bibinfo {author}
  {\bibfnamefont{A.~N.}\ \bibnamefont{Polyakov}}, \bibinfo {author}
  {\bibfnamefont{I.~V.}\ \bibnamefont{Shirokovsky}}, \bibinfo {author}
  {\bibfnamefont{Y.~S.}\ \bibnamefont{Tsyganov}}, \bibinfo {author}
  {\bibfnamefont{G.~G.}\ \bibnamefont{Gulbekian}}, \bibinfo {author}
  {\bibfnamefont{S.~L.}\ \bibnamefont{Bogomolov}}, \bibinfo {author}
  {\bibfnamefont{B.~N.}\ \bibnamefont{Gikal}}, \bibinfo {author}
  {\bibfnamefont{A.~N.}\ \bibnamefont{Mezentsev}}, \bibinfo {author}
  {\bibfnamefont{S.}~\bibnamefont{Iliev}}, \bibinfo {author}
  {\bibfnamefont{V.~G.}\ \bibnamefont{Subbotin}}, \bibinfo {author}
  {\bibfnamefont{A.~M.}\ \bibnamefont{Sukhov}}, \bibinfo {author}
  {\bibfnamefont{A.~A.}\ \bibnamefont{Voinov}}, \bibinfo {author}
  {\bibfnamefont{G.~V.}\ \bibnamefont{Buklanov}}, \bibinfo {author}
  {\bibfnamefont{K.}~\bibnamefont{Subotic}}, \bibinfo {author}
  {\bibfnamefont{V.~I.}\ \bibnamefont{Zagrebaev}}, \bibinfo {author}
  {\bibfnamefont{M.~G.}\ \bibnamefont{Itkis}}, \bibinfo {author}
  {\bibfnamefont{J.~B.}\ \bibnamefont{Patin}}, \bibinfo {author}
  {\bibfnamefont{K.~J.}\ \bibnamefont{Moody}}, \bibinfo {author}
  {\bibfnamefont{J.~F.}\ \bibnamefont{Wild}}, \bibinfo {author}
  {\bibfnamefont{M.~A.}\ \bibnamefont{Stoyer}}, \bibinfo {author}
  {\bibfnamefont{N.~J.}\ \bibnamefont{Stoyer}}, \bibinfo {author}
  {\bibfnamefont{D.~A.}\ \bibnamefont{Shaughnessy}}, \bibinfo {author}
  {\bibfnamefont{J.~M.}\ \bibnamefont{Kenneally}}, \bibinfo {author}
  {\bibfnamefont{P.~A.}\ \bibnamefont{Wilk}}, \bibinfo {author}
  {\bibfnamefont{R.~W.}\ \bibnamefont{Lougheed}}, \bibinfo {author}
  {\bibfnamefont{R.~I.}\ \bibnamefont{Il'kaev}},\ and\ \bibinfo {author}
  {\bibfnamefont{S.~P.}\ \bibnamefont{Vesnovskii}},\ }%
  \bibfield{journal}{%
  \Doi{10.1103/PhysRevC.70.064609}{\bibinfo {journal} {Phys. Rev. C}}\ }%
  \textbf{\bibinfo {volume} {70}},\ \bibinfo {pages} {064609} (\bibinfo {year}
  {2004})%
  \bibAnnoteFile{NoStop}{Oganessian04}%
\bibitem{Oganessian06}%
  \BibitemOpen
  \bibfield{author}{%
  \bibinfo {author} {\bibfnamefont{Y.~T.}\ \bibnamefont{Oganessian}},\ }%
  \bibfield{journal}{%
  \bibinfo {journal} {Phys. Scr.}\ }%
  \textbf{\bibinfo {volume} {T125}},\ \bibinfo {pages} {57} (\bibinfo {year}
  {2006})%
  \bibAnnoteFile{NoStop}{Oganessian06}%
\bibitem{Doukellis88}%
  \BibitemOpen
  \bibfield{author}{%
  \bibinfo {author} {\bibfnamefont{G.}~\bibnamefont{Doukellis}}, \bibinfo
  {author} {\bibfnamefont{G.}~\bibnamefont{Hlawatsch}}, \bibinfo {author}
  {\bibfnamefont{B.}~\bibnamefont{Kolb}}, \bibinfo {author}
  {\bibfnamefont{A.}~\bibnamefont{Miczaika}}, \bibinfo {author}
  {\bibfnamefont{G.}~\bibnamefont{Rosner}},\ and\ \bibinfo {author}
  {\bibfnamefont{B.}~\bibnamefont{Sedelmeyer}},\ }%
  \bibfield{journal}{%
  \Doi{DOI: 10.1016/0375-9474(88)90108-X}{\bibinfo {journal} {Nucl. Phys.}}\ }%
  \textbf{\bibinfo {volume} {A485}},\ \bibinfo {pages} {369 } (\bibinfo {year}
  {1988})%
  \bibAnnoteFile{NoStop}{Doukellis88}%
\bibitem{Winkler81}%
  \BibitemOpen
  \bibfield{author}{%
  \bibinfo {author} {\bibfnamefont{U.}~\bibnamefont{Winkler}}, \bibinfo
  {author} {\bibfnamefont{R.}~\bibnamefont{Giraud}}, \bibinfo {author}
  {\bibfnamefont{H.}~\bibnamefont{Gr{\"a}f}}, \bibinfo {author}
  {\bibfnamefont{A.}~\bibnamefont{Karbach}}, \bibinfo {author}
  {\bibfnamefont{R.}~\bibnamefont{Novotny}}, \bibinfo {author}
  {\bibfnamefont{D.}~\bibnamefont{Pelte}},\ and\ \bibinfo {author}
  {\bibfnamefont{G.}~\bibnamefont{Strauch}},\ }%
  \bibfield{journal}{%
  \Doi{DOI: 10.1016/0375-9474(81)90059-2}{\bibinfo {journal} {Nucl. Phys.}}\ }%
  \textbf{\bibinfo {volume} {A371}},\ \bibinfo {pages} {477 } (\bibinfo {year}
  {1981})%
  \bibAnnoteFile{NoStop}{Winkler81}%
\bibitem{Pelte81}%
  \BibitemOpen
  \bibfield{author}{%
  \bibinfo {author} {\bibfnamefont{D.}~\bibnamefont{Pelte}}, \bibinfo {author}
  {\bibfnamefont{U.}~\bibnamefont{Winkler}}, \bibinfo {author}
  {\bibfnamefont{R.}~\bibnamefont{Novotny}},\ and\ \bibinfo {author}
  {\bibfnamefont{H.}~\bibnamefont{Gr{\"a}f}},\ }%
  \bibfield{journal}{%
  \Doi{DOI: 10.1016/0375-9474(81)90058-0}{\bibinfo {journal} {Nucl. Phys.}}\ }%
  \textbf{\bibinfo {volume} {A371}},\ \bibinfo {pages} {454 } (\bibinfo {year}
  {1981})%
  \bibAnnoteFile{NoStop}{Pelte81}%
\bibitem{Manduchi85}%
  \BibitemOpen
  \bibfield{author}{%
  \bibinfo {author} {\bibfnamefont{C.}~\bibnamefont{Manduchi}}, \bibinfo
  {author} {\bibfnamefont{M.~T.}\ \bibnamefont{{Rucco-Manduchi}}}, \bibinfo
  {author} {\bibfnamefont{G.~F.}\ \bibnamefont{Segato}},\ and\ \bibinfo
  {author} {\bibfnamefont{F.}~\bibnamefont{Andolfato}},\ }%
  \bibfield{journal}{%
  \bibinfo {journal} {Il Nuovo Cim.}\ }%
  \textbf{\bibinfo {volume} {89}},\ \bibinfo {pages} {225} (\bibinfo {year}
  {1985})%
  \bibAnnoteFile{NoStop}{Manduchi85}%
\bibitem{Majka87}%
  \BibitemOpen
  \bibfield{author}{%
  \bibinfo {author} {\bibfnamefont{Z.}~\bibnamefont{Majka}}, \bibinfo {author}
  {\bibfnamefont{M.~E.}\ \bibnamefont{Brandan}}, \bibinfo {author}
  {\bibfnamefont{D.}~\bibnamefont{Fabris}}, \bibinfo {author}
  {\bibfnamefont{K.}~\bibnamefont{Hagel}}, \bibinfo {author}
  {\bibfnamefont{A.}~\bibnamefont{Menchaca-Rocha}}, \bibinfo {author}
  {\bibfnamefont{J.~B.}\ \bibnamefont{Natowitz}}, \bibinfo {author}
  {\bibfnamefont{G.}~\bibnamefont{Nebbia}}, \bibinfo {author}
  {\bibfnamefont{G.}~\bibnamefont{Prete}}, \bibinfo {author}
  {\bibfnamefont{B.}~\bibnamefont{Sterling}},\ and\ \bibinfo {author}
  {\bibfnamefont{G.}~\bibnamefont{Viesti}},\ }%
  \bibfield{journal}{%
  \Doi{10.1103/PhysRevC.35.2125}{\bibinfo {journal} {Phys. Rev. C}}\ }%
  \textbf{\bibinfo {volume} {35}},\ \bibinfo {pages} {2125} (\bibinfo {year}
  {1987})%
  \bibAnnoteFile{NoStop}{Majka87}%
\bibitem{Viesti88}%
  \BibitemOpen
  \bibfield{author}{%
  \bibinfo {author} {\bibfnamefont{G.}~\bibnamefont{Viesti}}, \bibinfo {author}
  {\bibfnamefont{B.}~\bibnamefont{Fornal}}, \bibinfo {author}
  {\bibfnamefont{D.}~\bibnamefont{Fabris}}, \bibinfo {author}
  {\bibfnamefont{K.}~\bibnamefont{Hagel}}, \bibinfo {author}
  {\bibfnamefont{J.~B.}\ \bibnamefont{Natowitz}}, \bibinfo {author}
  {\bibfnamefont{G.}~\bibnamefont{Nebbia}}, \bibinfo {author}
  {\bibfnamefont{G.}~\bibnamefont{Prete}},\ and\ \bibinfo {author}
  {\bibfnamefont{F.}~\bibnamefont{Trotti}},\ }%
  \bibfield{journal}{%
  \Doi{10.1103/PhysRevC.38.2640}{\bibinfo {journal} {Phys. Rev. C}}\ }%
  \textbf{\bibinfo {volume} {38}},\ \bibinfo {pages} {2640} (\bibinfo {year}
  {1988})%
  \bibAnnoteFile{NoStop}{Viesti88}%
\bibitem{Fornal90}%
  \BibitemOpen
  \bibfield{author}{%
  \bibinfo {author} {\bibfnamefont{B.}~\bibnamefont{Fornal}}, \bibinfo {author}
  {\bibfnamefont{F.}~\bibnamefont{Gramegna}}, \bibinfo {author}
  {\bibfnamefont{G.}~\bibnamefont{Prete}}, \bibinfo {author}
  {\bibfnamefont{G.}~\bibnamefont{Nebbia}}, \bibinfo {author}
  {\bibfnamefont{R.}~\bibnamefont{Smith}}, \bibinfo {author}
  {\bibfnamefont{G.}~\bibnamefont{D'Erasmo}}, \bibinfo {author}
  {\bibfnamefont{L.}~\bibnamefont{Fiore}}, \bibinfo {author}
  {\bibfnamefont{A.}~\bibnamefont{Pantaleo}}, \bibinfo {author}
  {\bibfnamefont{G.}~\bibnamefont{Viesti}}, \bibinfo {author}
  {\bibfnamefont{P.}~\bibnamefont{Blasi}}, \bibinfo {author}
  {\bibfnamefont{F.}~\bibnamefont{Lucarelli}}, \bibinfo {author}
  {\bibfnamefont{I.}~\bibnamefont{Iori}},\ and\ \bibinfo {author}
  {\bibfnamefont{A.}~\bibnamefont{Moroni}},\ }%
  \bibfield{journal}{%
  \Doi{10.1103/PhysRevC.41.127}{\bibinfo {journal} {Phys. Rev. C}}\ }%
  \textbf{\bibinfo {volume} {41}},\ \bibinfo {pages} {127} (\bibinfo {year}
  {1990})%
  \bibAnnoteFile{NoStop}{Fornal90}%
\end{thebibliography}
\end{document}